\documentclass[11pt,a4paper]{article}

\usepackage{amsmath,amsfonts,amssymb}
\usepackage{amscd}
\usepackage{hyperref}
\usepackage{color}
\usepackage{braket}
\usepackage{stmaryrd}
\usepackage{mathrsfs}
\usepackage{slashed}

\makeatletter
\renewcommand{\theequation}{\arabic{section}.\arabic{equation}}
\@addtoreset{equation}{section}

\newlength{\minitwocolumn}
\def\DGO{\slashed{\mathcal{D}}}
\newfont{\mmfrak}{eufm10 scaled 1200}
\newcommand{\mfrak}[1]{\mbox{\mmfrak #1}}

\def\be{\begin{equation}}
\def\ee{\end{equation}}
\def\bea{\begin{eqnarray}}
\def\eea{\end{eqnarray}}
\def\benu{\begin{enumerate}}
\def\eenu{\end{enumerate}}
\def\half{{1\over 2}}
\def\p{\partial}
\def\braket#1{\langle#1\rangle}
\def\definition#1{\medskip\noindent{\bf Definition\ } {\sl #1} :\ }
\def\proof{\medskip\noindent{\sl Proof\ } :\  }
\def\id{{\bf1}}
\def\Matrix#1#2{\left(\begin{array}{#1}#2\end{array}\right)}

\def\GL{{\mfrak L}}
\def\GA{{\mfrak A}}
\def\GJ{{\mfrak J}}
\def\Gs{{\mfrak s}}

\def\CR{{\cal R}}
\def\CL{{\cal L}}
\def\CJ{{\cal J}}
\def\CF{{\cal F}}
\def\CP{{\cal P}}
\def\CH{{\cal H}}

\def\BR{{\bf R}} 

\begin{document}
\begin{titlepage}
\begin{flushright}
\null \hfill Preprint TU-1101\\[3em]
\end{flushright}

\begin{center}
{\Large \bf
Metric algebroid and Dirac generating operator \\in Double Field Theory
}

\vskip 1.2cm

Ursula Carow-Watamura\footnote{E-mail:~ursula@tuhep.phys.tohoku.ac.jp}, 
Kohei Miura\footnote{E-mail:~miura@tuhep.phys.tohoku.ac.jp}, 
 Satoshi Watamura\footnote{E-mail:~watamura@tuhep.phys.tohoku.ac.jp},
and Taro Yano\footnote{E-mail:~taro.yano0115@gmail.com} 

\vskip 0.4cm
{
\it
Particle Theory and Cosmology Group, \\
Department of Physics, Graduate School of Science, \\
Tohoku University \\
Aoba-ku, Sendai 980-8578, Japan \\ 

\vskip 0.4cm
}

\vskip 0.4cm

\begin{abstract}
We give a formulation of Double Field Theory (DFT) based on a metric algebroid.  We derive a covariant completion of the Bianchi identities, i.e. the pre-Bianchi identity in torsion and an improved generalized curvature, and the pre-Bianchi identity  including the dilaton contribution. The derived bracket formulation by the Dirac generating operator is applied to the metric algebroid. 
We propose a generalized Lichnerowicz formula and show that it is equivalent to the pre-Bianchi identities. The dilaton in this setting is included as an ambiguity in the divergence. The projected generalized Lichnerowicz formula gives a new formulation of the DFT action. The closure of the generalized Lie derivative on the spin bundle yields the Bianchi identities as a consistency condition. A relation to the generalized supergravity equations (GSE) is discussed.

\end{abstract}


\medskip


\end{center}
\end{titlepage}

\newpage

\tableofcontents

\flushbottom
\setcounter{tocdepth}{3}

\section{Introduction}

Recently, algebroid structures are being explored with the aim to characterize the effective theories of string geometrically 
in frameworks such as the generalized geometry, double field theory and exceptional field theory.
In generalized geometry \cite{Hitchin:2005in,Gualtieri:aa}, we consider a generalization of the tangent vector, i.e., a generalized vector in  
$T^*M\oplus TM$  over a given manifold $M$, and an action of $O(D,D)$ as a rotation 
of the generalized vector. For a review see \cite{Hitchin_2011,Bouwknegt_2010}.
The formulation of the supergravity in generalized geometry setting 
is based on the structure of a Courant algebroid \cite{Coimbra_2011,Coimbra:2012aa}. 

Double field theory (DFT) has been developed with the aim to formulate a T-duality invariant, gauge invariant theory \cite{Hull:2009aa,Hohm_2010}.
The similarity between generalized geometry and DFT is well known and has been used to develop the theory from the early stage. 
See \cite{Zwiebach_2012,Aldazabal_2013} for a review and references therein. The main difference is that
in DFT the base manifold becomes twice the dimension of original manifold $M$, while 
in generalized geometry only the dimension of the fiber space is doubled.
The DFT picture is natural from the string point of view, since we consider the string moving
in the dual manifold $\tilde M$ after T-dual transformation.
Thus, DFT is defined on a 2D dimensional manifold $\mathbb{M}=\tilde M\times M$ and 
the generalized tangent vector is a section of the tangent bundle $T\mathbb{M}$.
However, the algebraic structure on $T\mathbb{M}$ is not the standard Lie algebra of tangent vectors but an algebroid,
 which reduces to a Courant algebroid when reducing the DFT to the supergravity frame.

In standard DFT, usually a differential constraint on the fields is imposed 
to obtain the D-dimensional theory, called the section condition, which is associated to the matching condition of the spectrum in string theory.
However, the section condition depends on the explicit choice of the local coordinates.
Moreover,
it is expected that non-geometric flux will rather be obtained by section-condition-violating configurations and, it is desirable to have the formulation based on the symmetry and independent of the section condition \cite{Dibitetto_2012,Aldazabal_2013,Geissbuhler:2013uka}.  
Of course, the dimension of the spacetime of the DFT is doubled and thus, 
eventually, we need to reduce the theory back to its original dimensional spacetime.  

Basic notions of DFT in differential-geometric terms have been proposed in \cite{Vaisman_2012} where the author considers the 2D dimensional manifold $\mathbb{M}$ as a flat, para-K\"ahler manifold with a Courant-like bracket defined on its tangent bundle,
which is called metric algebroid. 
The geometric aspects in DFT were also investigated in \cite{Chatzistavrakidis_2018,Marotta_2018,chatzistavrakidis2019algebroid}. 
See also \cite{Mori:2019aa}. 

There is also an approach to DFT using the generalization of a QP-manifold or differential graded manifold \cite{Roytenberg:1999} (see also \cite{Cattaneo_2001,Ikeda_2017} and reference therein). In ref.  \cite{Deser_2015}, it has been pointed out that the bracket structure of DFT can be obtained by using the differential graded manifold method. Then, the master equation was relaxed, which is called a pre-QP manifold, and the consistency condition was derived as a weak master equation \cite{Carow_Watamura_2016,Deser_2018}. When the master equation is relaxed, we are dealing with a metric algebroid.

In the pre-QP-manifold approach to DFT it is natural to analyze the Bianchi identities of the fluxes from the point of view of the weak master equation \cite{Carow_Watamura_2019}. The QP-manifold approach is a kind of BRST-BV approach and the master equation is related with the closure condition of the underlying algebroid. From this point of view, we can say that in our previous paper, we confirmed that the Bianchi identity of DFT can be obtained from the condition of closure on the metric algebroid. 
Furthermore, in this analysis we found a pre-Bianchi identity, which gives the consistency of the algebroid of DFT before imposing the weak master equation \cite{Carow_Watamura_2019}. 
 Besides being consistent with the standard DFT, this formulation can also include more structure on the 
base manifold, e.g. group manifolds, as discussed in \cite{Blumenhagen:2014gva,Hassler2016TheTO}. 

Recently, in the generalized geometry framework some developments to include the dilaton have been worked out in \cite{Garcia-Fernandez:2016aa,Severa:2018aa,JURCO20181}. One method is to use the divergence operation in a Courant algebroid \cite{AlekseevDGO} to characterize the dilaton in the framework of generalized geometry. The authors used the derived bracket formulation by the Dirac generating operator \cite{AlekseevDGO}. 
This formulation can be understood as a quantization of the graded Poisson structure of the QP manifold \cite{Grutzmann:2014aa}.

Our motivation in this paper is to apply the Dirac generating operator (DGO) formulation to DFT, which will provide us with a 
mechanism to include the dilaton into the theory. 
However, unlike in generalized geometry, we will not require the 
square of the DGO to be a function, which is the analog of the relaxation of the master equation in the pre-QP manifold approach. This strategy will lead us to a relation between DGO and the pre-Bianchi identities which we can use to characterize the class of metric algebroid underlying DFT. We give a 
generalized Lichnerowicz formula for DFT and show that it is equivalent to the condition that the pre-Bianchi identity is satisfied. 
From the projected Lichnerowicz formula we derive an action for DFT. From the closure condition of the generalized Lie derivative on the spin bundle we obtain the Bianchi identities including the dilaton contribution. 

The organization of this paper is the following: 

In section 2, we give a brief overview on Courant algebroid, metric algebroid and the Jacobi identities involved. 

In section 3, we formulate the metric algebroid underlying DFT and the base independent form of the relations of bracket and anchor is discussed.
Then, the generalized curvature tensor in this metric algebroid is constructed which enjoys tensorial properties. We derive the pre-Bianchi identity in curvature and torsion, and the pre-Bianchi identity for the dilaton. Formulae for rotation invariance of the frame are given. A generalization of the anchor map is also discussed. 

In section 4, the derived bracket formulation by Dirac generating operator and an explicit expression for this Dirac operator are given. Then, the requirement of the pre-Bianchi identity on the structure functions is reformulated as the statement that a generalized Lichnerowicz formula holds. 
The curvature scalar which appears in this generalized Lichnerowicz formula coincides with the scalar curvature obtained by taking the contraction of the generalized curvature tensor constructed in section 3.

In section 5, we introduce a Riemann structure by a splitting of the vector bundle. Then we define the Dirac operator 
compatible with the projection.
 The projected generalized Lichnerowicz formula is defined, and a proposal for a DFT action of the NS-NS sector is given. 

In section 6, closure properties and Bianchi identities are analyzed. 

In section 7, conclusions and discussions are given.  A connection to the generalized supergravity 
equations (GSE) via the structure function $F_A$ is proposed.

\section{Courant algebroid and metric algebroid structure}

Before we start discussing the metric algebroid, we briefly recall here the definition of a
Courant algebroid \cite{Courant_1990} for convenience. We follow \cite{Vaisman_2005}.

\subsection{Preliminaries: Courant algebroid} 

A Courant algebroid is a vector bundle $E$ over a base manifold $M$, endowed with  
a bracket $[-,-] : \Gamma(E)\times \Gamma(E)\rightarrow \Gamma(E)$ on the sections $\Gamma(E)$, 
 a bundle map (anchor) $\rho$ to the tangent bundle $TM$, 
$\rho: \Gamma(E)\rightarrow \Gamma(TM)$, and a non-degenerate symmetric fiber metric
$\braket{-,-} $, satisfying 
\begin{alignat}{3}
&(a)~~~~~&\rho(a)\braket{b,c}&=\braket{[a,b],c}
+\braket{b,[a,c]}~,
\\&(b)~~~~~&[a,a]&=\half\p\braket{a,a}~,
\\ &(c)~~~~~&[a,[b,c]]&=[[a,b],c]+[b,[a,c]]~, \label{LeibnizidentityofCA}
\end{alignat}
for $a,b,c\in\Gamma(E)$ and $f\in C^\infty(M)$,
where a differential $\p:C^\infty(M)\rightarrow \Gamma(E)$ is defined as $\braket{\p f,a}=\rho(a)f$.

The first equation is the compatibility of the bracket with the fiber metric.
The second property shows that the bracket is not necessarily anti-symmetric up to a derivative term.
The last identity is the Jacobi identity in form of the Leibniz rule of the bracket. 
Since the bracket is not necessarily anti-symmetric, it is not equivalent to a cyclic form of the Jacobi identity, in general.
Throughout this paper, Jacobi identity means the Jacobi identity of Leibniz form, unless we state differently.

From the above defining equations, various properties of the bracket can be derived. Important formulae are 
\begin{alignat}{3}
&(d)~~~~~&[a,fb]&=(\rho(a)f)b+f[a,b]~, \label{property d of CA}
\\&(e)~~~~~&[fa,b]&=-(\rho(b)f)a+(\p f)\braket{a,b}+f[a,b]~,\label{property e of CA}
\\&(f)~~~~~&[\p f,a]&=0~,\label{property f of CA}
\\&(g)~~~~~&\rho(\p f)&=0~,\label{property g of CA}
\\&(h)~~~~~&\rho([a,b])&=[\rho(a),\rho(b)]_L~,\label{pCompatibilityOfanchor1}
\end{alignat}
where the bracket $[-,-]_L$ is the standard Lie bracket on $TM$.

The identity (d) shows that the bracket is a derivation w.r.t. the second argument,
which follows from (a). 
(e) can be proven from (d) together with (b). 
The identities (f) and  (g) are the consequence of (e) and (h), while (h) itself is the consequence of the
Jacobi identity (c). The relations (f)-(h) will be discussed below.

\subsection{Metric algebroid}

In this paper, we use the following metric algebroid as the underlying symmetry structure of DFT.
We consider a
vector bundle $E\rightarrow{M}$. As in the Courant algebroid case,
we introduce a bracket $[-,-]:\Gamma(E)\times \Gamma(E)\rightarrow \Gamma(E)$,
an inner product $\braket{-,-}:\Gamma(E)\times \Gamma(E)\rightarrow C^\infty({M})$,
a bundle map (anchor) $\rho:E\rightarrow T{M}$, and a differential $\p:C^\infty({M})\rightarrow \Gamma(E)$ s.t. $\braket{\p f,a}=\rho(a)f$.

\definition{Metric algebroid \rm{\cite{Vaisman_2012}}} A vector bundle $(E,[-,-],\braket{-,-},\rho)$ is called metric algebroid if it satisfies
\begin{alignat}{3}
&(a)~~~~~&\rho(a)\braket{b,c}&=\braket{[a,b],c}
+\braket{b,[a,c]}~~,
\\&(b)~~~~~&[a,a]&=\half \p\braket{a,a}~~,\label{pCAaxiom2}
\end{alignat}
where $a,b,c\in\Gamma(E)$. Compared to the Courant algebroid, the Jacobi identity of the bracket is dropped.
Thus, it follows also that the compatibility of the anchor with the bracket (\ref{pCompatibilityOfanchor1}) does not hold, in general.
This bracket corresponds to the Dorfmann bracket in the Courant algebroid.

In order to discuss the correspondence with DFT, we need to introduce a set of local basis vectors $E_A$ on the bundle $E$, s.t.
\be
\braket{E_A,E_B}=\eta_{AB}~~,\label{pLocalinnerproduct}
\ee
where $\eta_{AB}$ is a symmetric constant tensor. We introduce $\eta^{AB}$ by
$\eta_{AB}\eta^{BC}=\delta_A^C$ and the raising and lowering of indices by $\eta$. 
In this basis we can write the differential operator as
\be
\p f=\sum_A (\rho(E_A)f) E^A~~.
\ee
We then define a structure function $F_{AB}{}^C\in C^\infty({M})$ of the bracket by
\be
[E_A,E_B]=F_{AB}{}^CE_C~~.
\ee
Using this basis, we can show that 
\be
F_{ABC}=\braket{[E_A,E_B],E_C}=F_{AB}{}^D\eta_{DC}
\ee
is totally antisymmetric.

\proof 
 For any $a,b\in \Gamma(E)$, 
 applying the second equation in the definition to $[a+b,a+b]$, we obtain
\be
[a,b]+[b,a]= \p\braket{a,b}~~,\label{pSymmetricpartofbracket pCA}
\ee
In the basis { $\braket{E_A,E_B}=\eta_{AB}=const$}, the r.h.s. is zero and therefore, $F_{ABC}=-F_{BAC}$. Furthermore, by the compatibility with the fiber metric we have 
\be
0=\rho(c)\braket{a,b}=\braket{[c,a],b}+\braket{a,[c,b]}
=\braket{[c,a],b}+\braket{[c,b],a}~~,
\ee
and therefore $F_{CAB}=-F_{CBA}$.
Combining the above two relations, we obtain $F_{ABC}=-F_{BAC}=F_{BCA}$,
i.e., $F_{ABC}$ is cyclic symmetric and thus totally anti-symmetric.

\subsection{Jacobi identity}
Since in the metric algebroid we do not require the Jacobi identity we define here a quantity which traces the deviation of the Jacobi identity from the Courant algebroid.
For this purpose we define the following maps $\GL:\Gamma(E)\times\Gamma(E)\times\Gamma(E)
\rightarrow\Gamma(E)$ and $\GL':\Gamma(E)\times\Gamma(E)\rightarrow 
\Gamma(TM)$:
\bea
\GL(a,b,c)&=&[a,[b,c]]-[[a,b],c]-[b,[a,c]] ~,  \label{pCAJacobiator}
\\ \GL'(a,b)&=&\rho([a,b])-[\rho(a),\rho(b)]_L ~,\label{pCAJacobiator2}
\eea
where $[-,-]_L$ denotes the standard Lie bracket on $TM$.
The map $\GL$ in (\ref{pCAJacobiator}) is a Jacobiator in Leibniz like form.
We added here $\GL'$ which does not vanish in general.

These quantities satisfy the following relations:
\bea 
\GL(a,b,c)+ \GL(b,a,c)&=&-[\p\braket{a,b},c]~, \label{pSymmetricpartofGLinpreCA}
\\ \GL'(a,b)+\GL'(b,a)&=&\rho(\p\braket{a,b})~. \label{psymmetricpartofGL'}
\eea
These relations follow from (\ref{pSymmetricpartofbracket pCA}).  

Note that these maps are not $C^\infty(M)$-linear in all arguments. Explicitly, one obtains
\bea
\Delta\GL(a,b,c)&:=&\GL(fa,b,c)+\GL(a,gb,c)+\GL(a,b,hc)-(f+g+h)\GL(a,b,c)
\cr&=&-(\GL'(b,c)f)a
+(\GL'(a,c)g)b
-(\GL'(a,b)h)c
\cr&&-\braket{a,b}[\p f,c]  
+\braket{a,c}[\p f,b]
-\braket{b,c}[\p g, a]~,
\eea
and
\bea
\Delta\GL'(a,b)&:=&
\GL'(fa,b)+\GL'(a,gb)-(f+g)\GL'(a,b)
\cr&=&\braket{a,b}\rho(\p f)~. \label{plinearityofLprimeinpreCA}
\eea
We can rewrite the above expressions in a more symmetric form by considering the following map:
\be
\phi(a,b,c,d)=\braket{\GL(a,b,c),d}~.\label{mapGLtoCM}
\ee
Then, the tensorial property is given by
\bea
\Delta \phi
&:=&\phi(fa,b,c,d)+\phi(a,gb,c,d)+\phi(a,b,hc,d)+\phi(a,b,c,kd)-(f+g+h+k)\phi(a,b,c,d)
\cr&=&-(\GL'(b,c)f)\braket{a,d}
+(\GL'(a,c)g)\braket{b,d}
-(\GL'(a,b)h)\braket{c,d}
\cr&&-\braket{a,b}  (\GL'(c,d)f)
+\braket{a,c}(\GL'(b,d)f)
-\braket{b,c}(\GL'(a,d)g)~.\label{ptensorialpropertyofphi}
\eea

The relations (d)-(h) given in (\ref{property d of CA})-(\ref{pCompatibilityOfanchor1}) which hold for a Courant algebroid 
should be reconsidered in the metric algebroid.
We see easily that the relations (d) and (e) in (\ref{property d of CA}) and (\ref{property e of CA}) hold also in the metric algebroid: Relation (d) can be proven by evaluating $\rho(e)\braket{fa,b}=\rho(e)(f\braket{a,b})$ in two ways as 
\be
\braket{[e,fa],b}+\braket{fa,[e,b]}=(\rho(e)f)\braket{a,b}+f(\rho(e)\braket{a,b})~,
\ee
and using axiom (a) on the l.h.s. 
From axiom (b) we have $[a,b]=-[b,a]+\p\braket{a,b}$. Applying this relation
 to the bracket on both sides of (d), we get the relation (e). 

In contrary, the relations (f)-(h) in (\ref{property f of CA})-(\ref{pCompatibilityOfanchor1}) do not hold in the metric algebroid in general. 
Concerning the relation (f) we obtain 
from (\ref{pSymmetricpartofGLinpreCA}), 
$[\p\braket{a,b},c]=-\GL(a,b,c)-\GL(b,a,c)$ which is not necessarily zero, and thus (f) does not necessarily hold.
Also the relation (g) 
is not kept, rather it is given by
(\ref{psymmetricpartofGL'}).

On the other hand, in the metric algebroid, there is a useful relation 
\be
\braket{[\p f,a],b}=\GL'(a,b)f~~.\label{p_usefulrelationinpreCA}
\ee
\proof By axiom (a), $\rho(a)\braket{\p f,b}=\braket{[a,\p f],b}+\braket{\p f,[a,b]}$
and therefore,
\bea
\braket{[\p f,a],b}
&=&\braket{\p\braket{\p f,a},b}-\braket{[a,\p f],b}
\cr&=&\rho(x)\rho(b)f-\rho(a)\rho(b) f+\rho([a,b])f
\cr&=&\GL'(a,x)f~.
\eea

To conclude, for the Courant algebroid $\GL=0$ and it follows also $\GL'f=0$, 
since $[\p f,a]=0$ in (\ref{p_usefulrelationinpreCA}). On the contrary, for the metric algebroid
(h) does not hold necessarily. It means that the metric algebroid reduces to a Courant algebroid if $\GL=0$. On the other hand, if we require $\GL\not=0$ but $\GL'f=0$, this defines a class of metric algebroid, called pre-Courant algebroid, discussed in \cite{Vaisman_2005, BRUCE2019254}.

\subsubsection{Jacobi identity on TM}

From the definition of $\GL$ and $\GL'$ 
and
using the Jacobi identity of the Lie bracket $[-,-]_L$ 
we obtain the following relation: 
\bea
\GJ(a,b,c)
&=&
\rho(\GL(a,b,c))-
\Big([\rho(a),\GL'(b,c)]_L
-[\GL'(a,b),\rho(c)]_L
-[\rho(b),\GL'(a,c)]_L
\cr&&+\GL'(a,[b,c])-\GL'([a,b],c)
-\GL'(b,[a,c])\Big)~,
\label{pTMJacobiconstraint}
\eea
where $\GJ(a,b,c): \Gamma(E)\times\Gamma(E)\times\Gamma(E)\rightarrow \Gamma(T{M})$  is the Jacobiator of the bracket $[-,-]_L$:
 \bea
 \GJ(a,b,c)=
 [\rho(a),[\rho(b),\rho(c)]_L]_L+[\rho(b),[\rho(c),\rho(a)]_L]_L+[\rho(c),[\rho(a),\rho(b)]_L]_L~.
 \label{pJacobiIdentityofLieBracket}
 \eea
 \proof The relation (\ref{pTMJacobiconstraint}) can be shown by taking the anchor of $\GL(a,b,c)$:
 \bea
\rho(\GL(a,b,c))
&=&\rho([a,[b,c]])-\rho([[a,b],c])-\rho([b,[a,c]])
\cr&=&
[\rho(a),[\rho(b),\rho(c)]_L]_L
-[[\rho(a),\rho(b)]_L,\rho(c)]_L
-[\rho(b),[\rho(a),\rho(c)]_L]_L)
\cr&&+[\rho(a),\GL'([b,c])]_L
+\GL'(a,[b,c])
-[\GL'([a,b]),\rho(c)]_L
\cr&&~~~~~~~~-\GL'([a,b],c)
-[\rho(b),\GL'(a,c)]_L)
-\GL'(b,[a,c])~.
\eea

 Since the bracket $[-,-]_L$ on $T{M}$ is the standard Lie bracket, the Jacobi identity holds and thus we obtain
 \be
 \GJ(a,b,c)=0~~.
 \ee
This means that the r.h.s. of (\ref{pTMJacobiconstraint}) also vanishes and is another relation between the
structure functions.

\subsubsection{$E$-connection}
On the metric algebroid $E$ we define an $E$-connection, $\nabla^E: \Gamma(E)\times\Gamma(E)\rightarrow\Gamma(E)$
compatible with the inner product $\braket{-,-}$.
The $E$-connection is defined by the standard relations: for $a,b,c\in\Gamma(E)$ and $f\in C^\infty({M})$
\bea
\nabla^E_{a}fb&=&(\rho(a)f)b+f\nabla^E_ab~,
\\\nabla^E_{fa}b&=&f\nabla^E_ab~.
\eea
We also require compatibility with the inner product:
\be
\rho(a)\braket{b,c}=\braket{\nabla^E_{a}b,c}+\braket{b,\nabla^E_{a}c}~.\label{pEconnectionCompatibility}
\ee
Using the basis $E_A$, the connection $\nabla^E$ is defined by  
\be
\nabla^E_{E_A}E_B=W_{AB}{}^CE_C~~,\label{Econnectioninlocalbasis}
\ee
where $W_{ABC}$ is a gauge field. In the following, we also use the abbreviation $\nabla_A^E=\nabla^E_{E_A}$ as long as it does not cause confusion.
Compatibility with the fiber metric yields that 
$W_{ABC}$ is antisymmetric in the last two indices:
\be
\rho(E_A)\braket{E_B,E_C}= \braket{W_{AB}{}^DE_D,E_C}+\braket{E_B,W_{AC}{}^DE_D}=W_{ABC}+W_{ACB}=0~.
\ee

\subsubsection{$E$-torsion}
Having defined the $E$-connection, one can introduce a corresponding $E$-torsion by
\be
T(a,b,c)=\braket{\nabla^E_{a}b-\nabla^E_{b}a-[a,b], c}+\braket{\nabla^E_{c}a,b}~.
\label{pEtorsionPreCA}
\ee
The three terms in the first bracket correspond to the definition of the standard torsion except that the bracket is now the (Dorfman type) bracket of the metric algebroid.
It is not $C^\infty(M)$-linear w.r.t. the first argument $a$, a property which is recovered by the last term \cite{AlekseevDGO,Gualtieri_2010}. The same torsion was also introduced in DFT context in \cite{Hohm_2013}.

Using the local basis, we obtain the $E$-torsion in the form including the structure function as:
\bea
T_{ABC}:= T(E_A,E_B,E_C)
=\half W_{[ABC]}-F_{ABC}~.\label{pEtorsioncomponents}
\eea
\proof
\bea
T_{ABC}
&=&T(E_A,E_B,E_C)=\braket{W_{ABD}E^D-W_{BAD}E^D-F_{ABD}E^D,E_C}
+\braket{W_{CAD}E^D,E_B}
\nonumber\\ &=&
\half W_{[ABC]}-F_{ABC}~.
\eea
From the definition of the torsion we see that given an E-connection $\nabla^E$, we can define a new connection $\nabla'^E$ as
\be
\nabla'^E_AE_B=W'_{AB}{}^CE_C=(W_{AB}{}^C-{1\over 3}T_{AB}{}^{C})E_C~~,\label{TorsionCancelledConnection}
\ee 
which defines a torsionless connection, since the torsion $T'$ of this new connection vanishes as
\be
T'_{ABC}=\half W'_{[ABC]}-F_{ABC}=\half W_{[ABC]}-T_{ABC}-F_{ABC}=0~~.
\ee
In other words, the connection $W'_{ABC}$ defined in (\ref{TorsionCancelledConnection}) is independent of the torsion of the original $E$-connection 
and defines an equivalence class of connections up to a totally anti-symmetric part
\cite{Garcia-Fernandez:2016aa}.

\section{The metric algebroid in DFT}
\subsection{Standard DFT}
In the standard DFT, we consider the $2D$-dimensional manifold $\mathbb{M}=M\times\tilde M$ and the tangent bundle $T\mathbb{M}$. 
While in generalized geometry a generalized vector is given by a section $v^m\p_m+\tilde v_m dx^m$ of the generalized tangent bundle $T^*M\oplus TM$,
 in DFT the corresponding generalized vector is a tangent vector in $T\mathbb{M}$, $V=v^m\p_m+\tilde v_m\tilde\p^m=V^M\p_M\in \Gamma(T\mathbb{M})$
 where the local coordinates are denoted by $x^M=(\tilde x_m,x^m)$ and 
the local basis on  $T\mathbb{M}$ is denoted by $\p_M=(\tilde \p^m,\p_m)$.
The inner product on $T\mathbb{M}$ is defined by adopting a natural contraction on $T^*M\oplus TM$, denoted as  
 $\braket{X,Y}=\braket{X,Y}_{TM}=\eta_{MN}X^MY^N$ for $X,Y\in\Gamma(T\mathbb{M})$, where $\eta_{MN}$ is a constant $O(D,D)$ metric.

The bracket of the standard DFT is the D-bracket defined for $X,Y\in T\mathbb{M}$ as
\be
[X,Y]_D=(X^M\p_MY^N-Y^M\p_MX^N+Y^M\p^NX_M)\p_N~,\label{pCAStandardDFTDbracket}
\ee
where indices are raised and lowered by $\eta_{MN}$ and $\eta^{MN}$.
It is easy to see that the D-bracket satisfies the axioms of a metric algebroid:
\bea
\braket{[X,Y]_D,Z}+\braket{Y,[X,Z]_D}
&=&
(X^M\p_MY^N)Z_N
+(X^M\p_MZ^N)Y_N=X^M\p_M\braket{Y,Z}~,
\eea
and
\be
[X,X]_D=\half (\p^N\braket{X,X})\p_N~.
\ee
Thus, standard DFT has the structure of  a metric algebroid ($T\mathbb{M}$,$[-,-]_D$,$\braket{-,-}$,$\rho$) with trivial anchor $\rho$
\cite{Vaisman_2012}.

\subsection{DFT condition}
In the present formulation, we consider a metric algebroid $(E, [-,-],\rho)$ 
of a vector bundle $E$ over $\mathbb{M}$ with local basis $E^A\in \Gamma(E)$.  
The bracket is characterized by the structure function $F_{AB}{}^C$.  The anchor $\rho:\Gamma(E)\rightarrow
 \Gamma(T\mathbb{M})$  is defined in this basis as $\rho(E_A)=E_A{}^M\p_M$. 
 The specific property of the present metric algebroid compared to a general metric algebroid is that the anchor is invertible. Note that in this paper we consider no internal symmetry, therefore $dim(E)=dim(T\mathbb{M})$.

On the tangent bundle we denote the inner product by $\braket{-,-}_{TM}$, and require for a vector field 
$a\in\Gamma(E)$ to satisfy
\be
\braket{\rho(a),\rho(b)}_{TM}=\braket{a,b}~~.
\ee
It means that $\eta_{AB}=E_A{}^ME_B{}^N\eta_{MN}$ where $\eta_{MN}=\braket{\p_M,\p_N}_{TM}$ is the metric on the base manifold which is 
not required to be constant. We also assume that $\eta_{AB}$ is an $O(D,D)$ metric so that the metric algebroid consistently includes the generalized geometry. 

To summarize, we are considering a specific metric algebroid satisfying
the following conditions:
\benu
\item The metric algebroid $E$ on the manifold $\mathbb{M}$ with $dim(E)=dim(T\mathbb{M})$.
\item The anchor $\rho(E_A)=E_A{}^M\p_M$ is invertible. 
\item There exists an inner product on $T\mathbb{M}$, s.t. $\braket{\rho(a),\rho(b)}_{TM}=\braket{a,b}$.
We also assume that for the local basis, $\eta_{AB}:=\braket{\rho(E_A),\rho(E_B)}_{TM}$ is an $O(D,D)$ metric.
\eenu
We call the above set of conditions the DFT condition. The standard DFT satisfies the DFT condition but additionally requires the vielbein $E_A{}^M$ to be an $O(D,D)$ element.

\subsection{Jacobi identity with DFT condition}

In the previous section, we introduced the maps $\GL$ and $\GL'$ in a general metric algebroid which trace the deviation from the Courant algebroid. Here, we discuss the Jacobi identities with DFT condition.
To make expressions more compact we consider the map $\phi:\Gamma(E)^{\times 4}\rightarrow C^\infty(\mathbb{M})$ introduced in (\ref{mapGLtoCM}), and the map $\phi':\Gamma(E)^{\times 3}\rightarrow C^\infty(\mathbb{M})$
\bea
\phi(a,b,c,d)&=&\braket{\GL(a,b,c),d}~,
\\\phi'(a,b,c)&=&\braket{\GL'(a,b),\rho(c)}_{TM}~. \label{pDefofphiprime}
\eea
We also introduce the structure functions corresponding to these maps by using the action on the local frame $E_A$:
\bea
\GL(E_A,E_B,E_C)&=&\phi_{ABC}{}^DE_D ~,
\\\GL'(E_A,E_B)&=&\phi'_{AB}{}^C\rho(E_C)~, \label{pLprimestructurefunction}
\eea
where they are represented by the above maps as
\be
\phi_{ABCD}=\phi(E_A,E_B,E_C,E_D)~~,~~~
\phi'_{ABC}=\phi'(E_A,E_B,E_C)~.
\ee
It is easy to see that the structure function $\phi_{ABCD}$ is
totally antisymmetric and represented by the structure function $F_{ABC}$ as:
\bea 
\phi_{ABCD}
&=&\braket{[E_A,[E_B, E_C]]-[[E_A, E_B],E_C] -[E_B, [E_A, E_C]],E_D}
\nonumber\\&=&{1\over4!}\Big(4\rho(E_{[A})F_{BCD]}
-3F_{[AB}{}^{A'}F_{CD]A'}\Big)~.\label{Leibnizidentitytensor}
\eea
From (\ref{psymmetricpartofGL'}) it is clear that the structure function $\phi'_{ABC}$
 is antisymmetric in the first two indices:
\be
\phi'_{ABC}=-\phi'_{BAC}~.
\ee

On the metric algebroid the function $\phi(a,b,c,d)$ is not $C^\infty(\mathbb{M})$-linear
w.r.t. all arguments, as we have seen in (\ref{ptensorialpropertyofphi}), nor is $\phi'(a,b,c)$. 
From (\ref{plinearityofLprimeinpreCA}) its transformation rule is obtained as:
\bea
\Delta \phi'(a,b,c)
&:=&\phi'(fa,b,c)+\phi'(a,gb,c)+\phi'(a,b,hc)-(f+g+h)\phi'(a,b,c)
\nonumber\\&=&\braket{a,b}\braket{\rho(\p f),\rho(c)}_{TM}~.
\eea
On the other hand, 
 the following map $\tilde\phi:\Gamma(E)^{\times4}\rightarrow C^\infty(M)$
is a tensor and totally antisymmetric: 
\be
\tilde \phi(a,b,c,d)=\phi(a,b,c,d)+\braket{\GL'(a,b),\GL'(c,d)}_{TM}+\braket{\GL'(a,d),\GL'(b,c)}_{TM}-\braket{\GL'(a,c),\GL'(b,d)}_{TM}~.\label{BreakingofLeibnizIdentity}
\ee
\proof
From (\ref{plinearityofLprimeinpreCA}), 
\bea
\Delta\braket{\GL'(a,b),\GL'(c,d)}_{TM}&:=& 
\braket{\GL'(fa,b),\GL'(c,d)}_{TM}+\braket{\GL'(a,gb),\GL'(c,d)}_{TM}
\cr&&+\braket{\GL'(a,b),\GL'(hc,d)}_{TM}
+\braket{\GL'(a,b),\GL'(c,kd)}_{TM}
\cr&&-(f+g+h+k)\braket{\GL'(a,b),\GL'(c,d)}_{TM}
\cr&=&\braket{a,b}\braket{\rho(\p f),\GL'(c,d)}_{TM}+\braket{c,d}\braket{\rho(\p h),\GL'(a,b)}_{TM}
\cr&=&\braket{a,b}(\GL'(c,d)f)+\braket{c,d}(\GL'(a,b)h)~.
\label{plinearityofLprimeinpreCAGLGL}
\eea
We can show that the transformation of the extra terms 
 cancel the $\Delta\phi$ in (\ref{ptensorialpropertyofphi}) and recover the tensorial property of $\tilde\phi$.

Now, we can prove the antisymmetry property of the map $\tilde\phi(a,b,c,d)$ 
by evaluation in the local basis:
\be
\tilde\phi(E_A,E_B,E_C,E_D)=\phi_{ABCD}+\phi'_{ABC'}\phi'_{CD}{}^{C'}+\phi'_{ADC'}\phi'_{BC}{}^{C'}-\phi'_{ACC'}\phi'_{BD}{}^{C'}~.
\label{pDefphitilde}
\ee
Since $\phi_{ABCD}$ is totally antisymmetric and $\phi'_{ABC}=-\phi'_{BAC}$, it is clear that the r.h.s. is totally antisymmetric 
and thus $\tilde\phi(a,b,c,d)$ is totally antisymmetric in all arguments.

While in a Courant algebroid, the condition $\phi_{ABCD}=0$ yields the Jacobi identity on the structure functions $F_{ABC}$, in the metric algebroid this condition is not covariant and depends on the local frame.
However, as we have seen above, in the metric algebroid the quantity $\tilde\phi_{ABCD}$ is a covariant tensor. Here, we use this property to characterize our metric algebroid by
\be
\tilde\phi(a,b,c,d)=0~.\label{preLeibnizIdentity}
\ee
This defines a class of metric algebroid which also includes the standard DFT.\footnote{When we reduce to standard DFT, $\phi'_{[AB}{}^E\phi'_{CD]E}=0$ due to the section condition, and the pre-Bianchi identity reduces to the Bianchi identity  $\phi_{ABCD}=0$ (\ref{preLeibnizIdentity}).}
It is also remarkable that in our previous analysis of DFT using the pre-QP-manifold, the local form of the condition (\ref{preLeibnizIdentity}), i.e. $\tilde\phi(E_A,E_B,E_C,E_D)=0$ was obtained as a pre-Bianchi identity \cite{Carow_Watamura_2019},
and we refer to the above condition as the pre-Bianchi identity, the justification of which will follow. 
In the local basis, the pre-Bianchi identity is expressed as
\bea
\tilde\phi(E_A.E_B,E_C,E_D)&=&
\phi_{ABCD}+{1\over 8}\phi'_{[AB}{}^E\phi'_{CD]E}
\cr&=&{1\over6}\rho(E_{[A})F_{BCD]}-{1\over8}F_{[AB}{}^EF_{CD]E}
+{1\over 8}\phi'_{[AB}{}^E\phi'_{CD]E}=0~. \label{explicitformofpreBianchi}
\eea

\subsubsection{Jacob identity on $T\mathbb{M}$}

As we have discussed, for a general metric algebroid we obtain an extra consistency condition 
from the Jacob identity of the Lie bracket on the tangent bundle $T\mathbb{M}$.
With DFT conditions it gives an additional identity for the structure functions.
To obtain this identity we evaluate (\ref{pTMJacobiconstraint}) 
in the local basis:
\bea
\GJ(E_A,E_B,E_C)
&=&\rho(\GL(E_A,E_B,E_C))
\cr&&-\Big(
[\rho(E_A),\GL'(E_B,E_C)]_L
-[\GL'(E_A,E_B),\rho(E_C)]_L
-[\rho(E_B),\GL'(E_A,E_C)]_L
\cr&&+\GL'(E_A,[E_B,E_C])-\GL'([E_A,E_B],E_C)
-\GL'(E_B,[E_A,E_C])\Big)~.
\eea
Since the Jacobi identity holds, $\GJ(a,b,c)=0$, by using the definition of the structure functions we obtain
\bea
\rho(\GL(E_A,E_B,E_C))
&=&
\half (\rho(E_{[A})\phi'_{BC]}{}^D)\rho(E_D)
+\half \phi'_{[AB}{}^{D'}(F-\phi')_{C]D'}{}^{D}\rho(E_{D})
\nonumber\\&&+\half F_{[BC}{}^{D'}\phi'_{A]D'}{}^{D}\rho(E_{D})
-\rho(\p F_{ABC})~.
\eea

Taking the inner product with $\rho(E_D)$ we obtain for $\phi_{ABCD}$ 
\bea
\phi_{ABCD}
&=&
\half (\rho(E_{[A})\phi'_{BC]D})
+\half \phi'_{[AB}{}^{D'}(F-\phi')_{C]D'D}
\cr&&+\half F_{[BC}{}^{D'}\phi'_{A]D'D}
-(\rho(E_D)F_{ABC})~~.
\label{pTMJacofstructurefunctions1}
\eea
 The same identity can be obtained by 
 using the local basis and the structure functions $F_{ABC}$ and $\phi'_{ABC}$.
 We define the structure function of the Lie bracket, a generalized geometric flux $F'_{ABC}$, as
\bea
 [\rho(E_A),\rho(E_B)]_L=(F_{AB}{}^C-\phi'_{AB}{}^C)\rho(E_C):= F'_{AB}{}^C\rho(E_C)~.
 \label{pDefofF'inpreCA}
 \eea
 Then, in the local basis we obtain the condition on this structure function $F'_{ABC}$ from $\GJ$ in (\ref{pJacobiIdentityofLieBracket})
as
 \bea
\GJ(E_A,E_B,E_C)&=&
\sum_\circlearrowright [\rho(E_A), [\rho(E_B),\rho(E_C)]_L]_L
\cr&=&\half\Big(
(\rho(E_{[A})F'_{BC]}{}^D)\rho(E_D)
+F'_{[BC}{}^DF'_{A]D}{}^{C'}\rho(E_{C'})\Big)~,
\eea
where $\circlearrowright$ means the sum over cyclic permutation of indices.
Thus, the tensor $\CJ_{ABCD}$ can be expressed by the structure function $F'_{ABC}$ as
\bea
\CJ_{ABCD}
&:=&\braket{\GJ(E_A,E_B,E_C),\rho(E_D)}_{TM}
\cr&=&\half(\rho(E_{[A})F'_{BC]D}
+F'_{[BC}{}^{C'}F'_{A]C'D})~.
\label{pJacobiIdentityofLieBracket1}
\eea
One can easily show that this is equivalent to (\ref{pTMJacofstructurefunctions1}) by substituting 
the definition of $F'_{ABC}$ in (\ref{pDefofF'inpreCA}).
\bea
2\CJ_{ABCD}
&=&\rho(E_{[A})(F-\phi')_{BC]D}
+(F-\phi')_{[BC}{}^{C'}(F-\phi')_{A]C'D}
\cr &=&
\rho(E_{[A})F_{BC]D}
-\rho(E_{[A})\phi'_{BC]D}
\cr&&+F_{[BC}{}^{C'}F_{A]C'D}
-F_{[BC}{}^{C'}\phi'_{A]C'D}
-\phi'_{[BC}{}^{C'}F_{A]C'D}
+\phi'_{[BC}{}^{C'}\phi'_{A]C'D}~.
\label{pJacobiIdentityofLieBracket2}
\eea
Thus, the Jacobi identity $\CJ_{ABCD}=0$ gives a condition on the structure functions.

\subsection{Generalized curvature on metric algebroid}

With DFT condition, we can still apply the same definition of the $E$-connection
and $E$-torsion as in a general metric algebroid. On the other hand, the curvature has to be
reconsidered.

\subsubsection{Generalized curvature in DFT}

We can define a curvature on a metric algebroid by
completing the $C^\infty(M)$-linearity of the standard definition of curvature,
similarly to the $E$-torsion.
We start from the generalized curvature introduced in DFT in \cite{Hohm_2013}:
\be
\CR^{HZ}(a,b,c,d)=\braket{(\nabla^E_a\nabla^E_b-\nabla^E_b\nabla^E_a)c-\nabla^E_{[a,b]}c,d}
+\half\braket{\nabla^E_{E_A}a,b}\braket{\nabla^E_{E^A}c,d}+(a,b\leftrightarrow c,d)~~.
\ee
For convenience, we introduce the quantity $R^\nabla(a,b,c,d)$:
\be
R^\nabla(a,b,c,d):=
\braket{(\nabla^E_a\nabla^E_b-\nabla^E_b\nabla^E_a)c-\nabla^E_{[a,b]}c,d}
+\half\braket{\nabla^E_{E_A}a,b}\braket{\nabla^E_{E^A}c,d}~.
\ee
Thus, 
\be
R^{HZ}(a,b,c,d)=R^\nabla(a,b,c,d)+R^\nabla(c,d,a,b)~.\label{splitofcurvature}
\ee
The tensorial property of $\CR^{HZ}$ is 
\bea
\Delta\CR^{HZ}(a,b,c,d)
&:=&\CR^{HZ}(fa,b,c,d)+\CR^{HZ}(a,gb,c,d)+\CR^{HZ}(a,b,hc,d)+\CR^{HZ}(a,b,c,kd)
\cr&&~~~~~~~~~~~~~~~~~~~~~~~~~~~~~-(f+g+h+k)\CR^{HZ}(a,b,c,d)
\cr&=&-(\GL'(a,b)h)\braket{c,d}-(\GL'(c,d)f)\braket{a,b}~,\label{ptransformationRHZ}
\eea
where in the last line $\GL'$ is the map defined in (\ref{pCAJacobiator2}). 
The above equation means that if the Jacobi identity holds, i.e., in Courant algebroid $\GL'h=0$ and in this case $\CR^{HZ}$ is a tensor.

On the other hand, in the metric algebroid we do not neglect the map $\GL'$, then the curvature $\CR^{HZ}$ does not have a tensorial property as shown above. However, with DFT condition we can consider the map 
$\braket{\GL'(a,b),\GL'(c,d)}_{TM}$. 
Comparing the tensorial property of $\CR^{HZ}$ given 
in (\ref{ptransformationRHZ}) and the transformation 
rule $\Delta\braket{\GL'(a,b),\GL'(c,d)}_{TM}$ in (\ref{plinearityofLprimeinpreCAGLGL}),
it is easy to see that the following combination is in fact $C^\infty(\mathbb{M})$-linear in all
arguments
\bea
\CR(a,b,c,d)=\CR^{HZ}(a,b,c,d)+\braket{\GL'(a,b),\GL'(c,d)}_{TM}~.\label{pCurvatureinprepreCA}
\eea
In the following, we refer to $\CR: \Gamma(E)^{\times 4}\rightarrow C^\infty(\mathbb{M})$ given above as the generalized curvature in the metric algebroid.
In the local basis $E_A$, the explicit form of the generalized curvature tensor $\CR_{ABCD}$ is
\bea
\CR_{ABCD}&:=&\CR(E_A,E_B,E_C,E_D)
\cr&=&R^\nabla_{ABCD}+R^\nabla_{CDAB}+\phi'_{ABE}\phi'_{CD}{}^E~,
\label{generalizedcurvatureinlocalbasis}
\eea
where
\be
R^\nabla_{ABCD}:=\rho(E_{[A})W_{B]CD}-W_{[A|C}{}^{E'}W_{|B]E'D}-F_{AB}{}^E W_{ECD}+\half W_{EAB}W^E{}_{CD}~.
\label{curvatureinlocalbasis}
\ee

\subsubsection{Pre-Bianchi identity and curvature}

Now, we are ready to discuss the covariant form of the Bianchi identity.
In standard DFT there is a Bianchi identity given in terms of curvature $\CR^{HZ}$ and torsion \cite{Hohm_2012,Hohm_2013}.
Furthermore, in our discussion using the supermanifold approach \cite{Carow_Watamura_2019} we encountered a corresponding pre-Bianchi identity. 
Motivated by this, in the following we discuss the covariance of these identities 
as a structure in the metric algebroid. 

We have defined a tensor $\tilde\phi$ and formulated the pre-Bianchi identity (\ref{preLeibnizIdentity})
which characterize the metric algebroid for DFT.
We give here the pre-Bianchi identity in terms of the generalized curvature $\CR$ defined above.
This is achieved by realizing that the following identity holds:
\be
3\GA\CR(a,b,c,d)-\tilde \phi(a,b,c,d)=\GA\Big(4(\nabla^E_aT)(b,c,d)+3\sum_A T(a,b,E_A)T(c,d,E^A)\Big)~,\label{pCovariantBianchiinpreCA}
\ee 
where $\nabla^E_a$ is a connection on $\otimes^3\Gamma(E)^*$ by the Leibniz rule and $\GA$ is an antisymmetrization map which defines a totally antisymmetric tensor for a map $A(a_1,a_2,\cdots,a_n)$ as
\be
\GA A(a_1,a_2,a_3,\cdots a_n)={1\over n!}\sum_\sigma sign(\sigma)\phi(a_{\sigma(1)},a_{\sigma(2)},a_{\sigma(3)},\cdots a_{\sigma(n)})~,
\ee
where $\sigma$ is a permutation.

The proof of (\ref{pCovariantBianchiinpreCA}) can be given by using a local basis $E_A$, in which we have defined $\phi_{ABCD}$ in (\ref{Leibnizidentitytensor}).
Using the explicit form of (\ref{curvatureinlocalbasis}) (in local basis) and (\ref{splitofcurvature}), 
we can show an identity which is similar to the one given in \cite{Hohm_2013}:  
\bea
\Big(4\nabla_{E_{[A}}T_{BCD]}+3T_{[AB}{}^ET_{CD]E}\Big)
&=&3\CR^{HZ}_{[ABCD]}-\Big(4\rho({E_{[A}})F_{BCD]}-3F_{[AB}{}^EF_{CD]E}\Big)~,
\label{pDirectcalcidentity}
\eea
where we have replaced the $E$-torsion on the l.h.s. by the connection and the structure function using (\ref{pEtorsioncomponents}).

As we have seen in (\ref{ptransformationRHZ}),
$\CR^{HZ}$ on the r.h.s is not a tensor in the metric algebroid. Therefore, we use (\ref{pCurvatureinprepreCA})
 to replace $\CR^{HZ}$ with the generalized curvature $\CR$ including a correction term. 
 Then, we see that by total antisymmetrization, this correction term 
 combined with the other terms on the r.h.s. of (\ref{pDirectcalcidentity}) exactly  produces the tensor $\tilde\phi_{ABCD}$.
Thus, we get the identity (\ref{pCovariantBianchiinpreCA}) in the 
 local basis. Since in the resulting expression each term is a tensor, we get the general form of the identity
 (\ref{pCovariantBianchiinpreCA}) which is independent of the choice of the frame.
 
Now, imposing the pre-Bianchi identity $\tilde\phi(a,b,c,d)=0$ we obtain a frame independent formula
\be
3\GA\CR(a,b,c,d)=\GA\Big(4(\nabla^E_aT)(b,c,d)+3\sum_A T(a,b,E_A)T(c,d,E^A)\Big)~,\label{ppreBianchiinpreCA}
\ee 
which is a pre-Bianchi identity in curvature and torsion. As we saw in \cite{Carow_Watamura_2019}, the pre-Bianchi identity is the {equation} which holds when the flux is given by the vielbein as in standard DFT, that is, the standard DFT parametrization 
by the generalized vielbein is a solution of this pre-Bianchi identity.

Note that the identity of the form
\be
3\GA\CR^{HZ}(a,b,c,d)-\GA \phi(a,b,c,d)=\GA\Big(4(\nabla^E_aT)(b,c,d)+3\sum_A T(a,b,E_A)T(c,d,E^A)\Big)
\label{pNonCovariantBianchiinpreCA}
\ee 
also holds. It means that this particular combination of the maps $\CR^{HZ}$ and $\phi$ is an element of $\otimes^4 \Gamma(E)^*$, i.e., is $C^\infty(\mathbb{M})$-linear in all arguments,
 although each term on the l.h.s separately does not have this property.

\subsubsection{Pre-Bianchi identity and dilaton}

In DFT, there is another type of Bianchi identity which includes the contribution of the dilaton. 
We will discuss the property of the dilaton field using the divergence operator in the next section.
Here, we focus on how the dilaton can be accommodated into this algebraic structure.  

We start with the tensor $\CJ_{ABCD}$ in (\ref{pJacobiIdentityofLieBracket2}).
Taking a trace w.r.t. the last two indices we obtain 
 the following tensor: 
\bea
\CJ_{ABC}{}^C&=&\braket{\GJ(E_A,E_B,E_C),\rho(E^C)}_{TM}
\cr&=&\rho(E_{C})F_{AB}{}^C
-\rho(E_{C})\phi'_{AB}{}^C
-\rho(E_{[A})\phi'_{B]C}{}^C
\cr&&-F_{AB}{}^{C'}\phi'_{CC'}{}^C
+\phi'_{AB}{}^{C'}\phi'_{CC'}{}^C~.
\label{pJacobionTM}
\eea

To see the relation to the Bianchi identity including the dilaton discussed in 
standard DFT, 
we add a vector $U_A$ which
satisfies
\be
\rho(E_{[A})U_{B]}-F_{AB}{}^CU_C+\phi'_{AB}{}^CU_C=0~. \label{pGSEgeneralizedvector}
\ee
As we see there is a non-trivial solution for this condition.

Now we can add this combination of $U_A$ to (\ref{pJacobionTM})
and define a flux with one index by 
\be
F_A=\phi'_{CA}{}^C+U_A~.
\ee
 Using this flux $F_A$ we can rewrite the above identity as 
 \be
{{\rho(E_{C})F_{AB}{}^C}}
-{{F_{AB}{}^{C'}F_{C'}}}
+{{\rho(E_{[A})(F_{B]}})}
=\rho(E_{C})\phi'_{AB}{}^C-\phi'_{AB}{}^{C'}F_{C'}~.
\label{pBianchiofOneIndex1}
\ee

This identity is equivalent to the one including the dilaton in standard DFT \cite{Aldazabal_2013,Geissbuhler:2013uka}.
In order to see this, we show that $U_A=2\rho(E_A) d$ satisfies above condition (\ref{pGSEgeneralizedvector}). Since
\be
\rho(E_{[A})\rho(E_{B]})d-F_{AB}{}^C\rho(E_C)d+\phi'_{AB}{}^C\rho(E_C)d=0~,
\ee
the flux given as
\be
F_A=\phi'_{CA}{}^C+2\rho(E_A)d\label{pSingleStructurefunctionindilaton}
\ee 
satisfies the pre-Bianchi identity (\ref{pBianchiofOneIndex1}).
We postpone the discussion of the identification of $U_A$ as an ambiguity in the divergence and 
the relation between the pre-Bianchi identity for $F_A$ and the algebraic structure to the next section.

For reduction to standard DFT where the field $d$ is identified with the dilaton and $\phi'$ reduces to the Weizenb\"ock connection,
 we can show that the r.h.s. of (\ref{pBianchiofOneIndex1}) can be written as
\be
\rho(E_{C})\phi'_{AB}{}^C-\phi'_{AB}{}^{C'}F_{C'} =
\half (\p^M\p_M E_{[A}{}^P)E_{B]P}
-2\Omega^C{}_{AB}\rho(E_C)d~,
\ee
which coincides with the formula (1.5) in \cite{Geissbuhler:2013uka}.

\subsubsection{Rotation invariance of the frame}
In this section, we check explicitly the
local Lorentz covariance of the above structure functions, i.e., the covariance under the frame rotation,
although it is rather apparent due to their tensorial structure. 
Thus, we consider the rotational group with dimension $dim(E)$, and we denote an infinitesimal transformation 
using the basis $E_A$ as
\be
\delta_{\Lambda}E_A=\Lambda_{A}{}^BE_B~.
\ee
Then, the invariance of the inner product imposes $\Lambda_{AB}+\Lambda_{BA}=0$, and the rotation is $O(D,D)$.

The tensorial property of the structure function $F_{ABC}$ can be derived 
from the following relations:
\bea
\Delta\braket{[a,b],c}
&:=&\braket{[fa,b],c}+\braket{[a,gb],c}+\braket{[a,b],hc}-
(f+g+h)\braket{[a,b],c}
\cr &=&\braket{a,b}\rho(c)f-\braket{a,c}\rho(b)f
+\braket{b,c}\rho(a)g~.
\eea
Therefore, evaluating in the basis, we get
\be
\Delta F_{ABC}:=\Delta\braket{[E_A,E_B],E_C}=\eta_{AB}\rho(E_C)f-\eta_{CA}\rho(E_B)f+\eta_{BC}\rho
(E_A)g~.
\ee
From this relation of the $C^{\infty}(\mathbb{M})$-linearity, we obtain the
transformation of the structure function $F_{ABC}$ by identifying the functions $f$ and $g$ with the transformation parameter $\Lambda_{AB}$ as
\be
\delta_\Lambda F_{ABC}
=\Lambda\triangleright F_{ABC}+\half \rho(E_{[A})\Lambda_{BC]}~,\label{pRotationOfF}
\ee
where $\Lambda\triangleright$ means the linear term of the transformation, i.e., $\Lambda\triangleright F_{ABC}$ is
\bea
\Lambda\triangleright F_{ABC}=\Lambda_A{}^{A'}F_{A'BC}+\Lambda_B{}^{B'}F_{AB'C}+\Lambda_C{}^{C'}F_{ABC'}~.
\eea

Similarly, we can get the transformation rules of the other quantities. 
In the following we list the transformations of the structure functions $\phi_{ABCD},\phi'_{ABC}$ 
and the maps $\GL(E_A,E_B,E_C)$, $\GL'(E_A,E_B)$ for convenience:
\bea
\delta_\Lambda \GL(E_A,E_B,E_C)&=&\Lambda\triangleright\GL(E_A,E_B,E_C)-\frac{1}{2}(\GL'(E_{[A},E_B)\Lambda_{C]}{}^{D})E_{D}-\frac{1}{2}[\partial\Lambda_{[AB},E_{C]}]~,
\\
\delta_\Lambda\phi_{ABCD}
&=&\Lambda\triangleright\phi_{ABCD}-\frac{1}{2}(\GL'(E_{[A},E_B)\Lambda_{C]D})+\frac{1}{2}\GL'(E_D,E_{[A})\Lambda_{BC]}~,
\\
\delta_\Lambda\GL'(E_A,E_B)
&=&\Lambda\triangleright\GL'(E_A,E_B)+\rho(\p\Lambda_{AB})~,
\\
\delta_\Lambda\phi'_{ABC}&=&\Lambda\triangleright \phi'_{ABC}+\rho(E_C)\Lambda_{AB}~.
\eea
For the connection $W_{ABC}$ in (\ref{Econnectioninlocalbasis})
\bea
\delta_\Lambda W_{ABC}&=&\braket{\nabla^E_{\Lambda_A{}^{A'}E_{A'}}E_B,E_C}+\braket{\nabla^E_{E_A}(\Lambda_B{}^{B'}E_{B'}),E_C}+\braket{\nabla^E_{E_A}E_B,\Lambda_C{}^{C'}E_{C'}}
\cr&=&\Lambda\triangleright W_{ABC}+\rho(E_{A})\Lambda_{BC}~,
\eea
as is required for the spin connection under local Lorentz transformation.
Note that the $\phi'_{BCA}$ has the same transformation property as the connection $W_{ABC}$.

\subsection{Generalized anchor map}

In standard DFT, the $D$-bracket is defined as (\ref{pCAStandardDFTDbracket}), i.e., directly
on the generalized vector $X^M\p_M\in\Gamma(T\mathbb{M})$.
This means that $T\mathbb{M}$ is identified with the metric algebroid.
 In the present formulation, we work with a metric algebroid on the vector bundle
$E$. From this point of view, the metric algebroid $T\mathbb{M}$ is 
  realized by the identification of the basis $E_A$ of the vector bundle $E$
with a generalized vector of $T\mathbb{M}$.
For this it is convenient to introduce a generalized anchor map,
which is a metric algebroid homomorphism
$\varphi:(E,\rho,[-,-],\braket{-,-})\rightarrow (T\mathbb{M},\rho_\varphi,[-,-]_\varphi,\braket{-,-}_\varphi)$
: for a vactor $a,b\in\Gamma(E)$,
\be
\varphi(a)=\rho(a)\in \Gamma(T\mathbb{M})~,
\ee
meaning that the basis is mapped as
\be
\varphi(E_A)=E_A{}^M\p_M\in \Gamma(T\mathbb{M})~,
\ee
the anchor becomes trivial, i.e.,
\be
\rho_\varphi=id~,
\ee
and the inner product is
\be
\varphi(\braket{a,b})=\braket{\varphi(a),\varphi(b)}_\varphi=\braket{\varphi(a),\varphi(b)}_{TM}~.
\ee
The bracket on the metric algebroid $E$ is mapped to $T\mathbb{M}$ 
as
\be
\varphi([a,b])=[\varphi(a),\varphi(b)]_\varphi~.
\ee
By using the relation for the bracket $[-,-]_\varphi$, we can evaluate the above relation in the basis:
\bea
\varphi([E_A,E_B])
&=&[\varphi(E_A),\varphi(E_B)]_\varphi
\cr&=&[E_A{}^M\p_M,E_B{}^N\p_N]_\varphi
\cr&=&
(E_{[A}{}^M\p_ME_{B]}{}^N)\p_N+E_B{}^N\eta_{MN}(\p E_A{}^M)+
E_A{}^ME_B{}^N[\p_M,\p_N]_\varphi~.
\eea
Then, we obtain that the original structure function $F_{ABC}$ can be expressed as
\bea
\braket{\varphi([E_A,E_B]),\varphi(E_C)}_\varphi
&=&F_{ABC}
\cr&=&\CF_{ABC}+
E_A{}^ME_B{}^NE_C{}^L\braket{[\p_M,\p_N]_\varphi,\varphi(\p_L)}_\varphi~,
\eea
where the function $\CF_{ABC}$ is given by
\be
\CF_{ABC}:= 
\Omega_{[AB]C}+\Omega_{CAB}~.\label{pStandardDFTflux}
\ee
We also introduce an affine connection 
$\nabla^{T\mathbb{M}}$ by
\be
\nabla^{T\mathbb{M}}_{\p_M}\p_N=\Gamma_{MN}{}^L\p_L~,
\ee
by employing the vielbein postulate as
\be
(\Omega_{AB}{}^C+\Gamma_{AB}{}^C-W_{AB}{}^C)E_C{}^N=0~,
\ee
where $\Gamma_{AB}{}^C=E_A{}^ME_B{}^NE^C{}_L\Gamma_{MN}{}^L$.

\paragraph{Reduction to standard DFT:}
The simplest case for this map $\varphi$ is that the structure function of the bracket $\braket{[\p_M,\p_N]_\varphi,\p_L}_\varphi=0$ which implies $\eta_{MN}$ is constant.
For this we obtain 
\be
F_{ABC}=\CF_{ABC}~,
\ee
and $\CF_{ABC}$ is totally antisymmetric,
which is the generalized flux in standard DFT. Then the  bracket of the tangent vectors $X,Y\in\Gamma(T\mathbb{M})$
is equivalent to the standard $D$-bracket  in (\ref{pCAStandardDFTDbracket}):
\be
[X,Y]_\varphi=[X,Y]_D~.
\ee
In the standard DFT, we require that $\eta^{MN}$ is an $O(D,D)$ metric and thus the vielbein
$E_A{}^M$ is also an element of $O(D,D)$.

\section{Derived bracket}

In a Courant algebroid, the bracket can be represented by a derived bracket \cite{Kosmann-Schwarzbach:2004aa}. 
Here, we want to apply this formulation to the metric algebroid of DFT. 
Motivated by \cite{AlekseevDGO,Garcia-Fernandez:2016aa,Severa_2017}, we define the bracket as a derived bracket on a Clifford bundle $Cl(E)$. Using the fiber metric, we can define a Clifford algebra
by introducing a product $\Gamma(Cl(E))\times \Gamma(Cl(E))\ni(a, b)\rightarrow  a  b\in\Gamma(Cl(E))$
with anti-commutation relation among the elements $a,b\in \Gamma(E)\subset\Gamma(Cl(E))$ as
\footnote{In principle, we have to distinguish an element $a\in \Gamma(E)$ and its Clifford action on
a Clifford module $\gamma(a)$ where $\gamma:\Gamma(E)\rightarrow \Gamma(Cl(E))$. 
To simplify the notation, we identify $\Gamma(E)$ and $\gamma(\Gamma(E))$ and do not write this action explicitly.}
\be
\{a,b\}=ab+ba=2\braket{a,b}~. \label{pCliffordbracketonframe}
\ee
We consider a connection $\nabla^{Cl} :  \Gamma(E)\times\Gamma(Cl(E))\rightarrow \Gamma(Cl(E))$ 
on the Clifford bundle induced by a given E-connection $\nabla^E$:  
On a section, $b\in\Gamma(E)\subset\Gamma(Cl(E))$ is defined by
\be
\nabla^{Cl}_ab=\nabla_a^Eb~, \label{pdef:InducedconnectiononCL}
\ee
and imposing Leibniz rule w.r.t. the Clifford product. Then,  
the compatibility with the fiber metric holds:
\be
\rho(a)\braket{b,c}=\braket{\nabla^{Cl}_ab,c}+\braket{b,\nabla^{Cl}_ac}~~.
\ee
The Clifford algebra can be considered as a quantization of a graded symplectic manifold or, equivalently, a QP-manifold \cite{Grutzmann:2014aa}.
In \cite{Carow_Watamura_2019}, we have shown that certain algebraic relations in DFT can be
 formulated on a pre-QP-manifold. Here, instead of the graded Poisson bracket of the pre-QP-manifold, we now consider the
graded commutator on the Clifford bundle and investigate the algebraic relations in metric algebroid formulation of DFT.
 
We consider a natural grading on the Clifford bundle 
where a section $\Gamma(E)\subset\Gamma(Cl(E))$ is odd and the bracket $\{-,-\}$ in (\ref{pCliffordbracketonframe})
is extended to $a,b\in Cl(E)$ as a graded bracket: 
\be
\{a, b\}=ab-(-1)^{| a|| b|}ba=-(-1)^{| a|| b|}\{ b, a\}~.\label{pDef:Gradedbracket}
\ee 
The degree of the element $a\in Cl(E)$ is denoted by $|a|$.
The bracket satisfies the graded Jacobi identity
which can be written in Leibniz form as
\be
\{a,\{b,c\}\}=\{\{a,b\},c\}+(-1)^{|a||b|}\{b,\{a,c\}\}~~.
\ee
A spin bundle $\mathbb{S}$ is a module over the Clifford bundle.
We introduce a connection on $\mathbb{S}$, 
\be
\nabla^{\mathbb{S}}~:~\Gamma(E)\times \Gamma(\mathbb{S})~\rightarrow~ \Gamma(\mathbb{S})~,
\ee 
for $e\in \Gamma(E)$, $\chi\in\Gamma(\mathbb{S})$ and $f\in C^\infty(M)$. It satisfies the standard property of a connection as
\be
\nabla_e^{\mathbb{S}}f\chi=(\rho(e)f)\chi+f\nabla_e^{\mathbb{S}}\chi
~~,~~~
\nabla_{fe}^{\mathbb{S}}\chi=f\nabla_e^{\mathbb{S}}\chi~.
\ee

We require the compatibility of this connection with the $E$-connection,
which means that the commutator with the spin connection
is defined by the connection on $Cl(E)$, see (\ref{pdef:InducedconnectiononCL}) :
for an element 
$a\in\Gamma(Cl(E))$ and $e\in\Gamma(E)$, a compatible connection on $\Gamma(\mathbb{S})$ satisfies 
\bea
\{\nabla^{\mathbb{S}}_{e}, a\}&=&\nabla^{Cl}_{e}a~,\label{pdef:inducedconnectiononS}
\eea
and for $f\in C^\infty(\mathbb{M})$
\be
\{\nabla^{\mathbb{S}}_{e}, f\}=\rho(e)f~.\label{pdef:inducedconnectiononSf}
\ee
It follows that for an element $a\in \Gamma(Cl(E))$  
\be
\nabla_{e}^{\mathbb{S}}a\chi=(\nabla_{e}^{Cl}a)\chi+a\nabla_e^{\mathbb{S}}\chi~.\label{pLeibnizofspinconnection}
\ee
Since the degree of $\nabla_a^{\mathbb{S}}$ is even and the
graded bracket in (\ref{pdef:inducedconnectiononS}) is an ordinary commutator,  
 (\ref{pLeibnizofspinconnection}) is a consequence of the Leibniz rule of the commutator.
Now, we are ready to introduce an odd differential operator, i.e., the Dirac operator $\slashed\DGO$.

\subsection{Derived bracket by Dirac generating operator}
\subsubsection{Derived bracket}

Given a metric algebroid $(E,\rho,[-,-],\braket{-,-})$ and
a graded commutator (\ref{pDef:Gradedbracket})
where the grading for an element in $\Gamma(Cl(E))$
is defined by the order of the element in $\Gamma(E)$.
This graded commutator has information of the inner product (\ref{pCliffordbracketonframe}).
On the Clifford bundle, we can construct
a differential operator $\DGO:\Gamma(\mathbb{S})\rightarrow\Gamma(\mathbb{S})$
 which generates all the structures of the metric algebroid, i.e.,
 $^\exists \DGO:\Gamma(\mathbb{S})\rightarrow\Gamma(\mathbb{S})$ generates
 the derivation $\p$, the bracket of the metric algebroid $[-,-]$ and
 the anchor map $\rho$ as follows. For $f\in C^\infty(\mathbb{M})$ and $a,b\in \Gamma(E)\subset\Gamma(Cl(E))$ 
\be
\p f=2\{\DGO,f\}~, \label{pDerivationforCA1}
\ee
\be
[a,b]=\{\{\DGO,a\},b\}~, \label{pDerivedBracketforCA1}
\ee
\be
\rho(a)f=\{\{\DGO,a\},f\}~. \label{pAnchorforCA1}
\ee
The bracket (\ref{pDerivedBracketforCA1}) generated by $\DGO$ is called a derived bracket.
From these relations, $\DGO$ is an odd graded linear differential operator, which is called Dirac generating operator \cite{AlekseevDGO}.
The concrete form of the Dirac generating operator $\DGO$ is discussed in \S\ref{section_DGO}.
The axioms of the metric algebroid can be 
derived using the above definitions and the Jacobi identity for the graded commutator:
\begin{alignat}{3}
a)&\quad 
\rho(a)\braket{b,c}&=&\braket{[a,b],c}+\braket{b,[a,c]}~, \label{pMetricanchorconsistencypreCA}
\\
b)&\quad 
\half\p\braket{a,a}&=&[a,a]~. \label{psymmetricpartDBpreCA}
\end{alignat}
\proof 
$a)$ follows from
\be
\{\{\DGO,a\},\{b,c\}\}=\{\{\{\DGO,a\},b\},c\}+\{b,\{\{\DGO,a\},c\}\}~,
\ee
$b)$ follows from
\be
\{\DGO,\{a,a\}\}=2\{\{\DGO,a\},a\}~.
\ee

As already mentioned, in the metric algebroid the Jacobi identity of the bracket is missing compared to the Courant algebroid.
We discuss here the breaking of the Jacobi identity in terms of the derived bracket. 
The Jacobi identity for Courant algebroid is usually derived by using the following relation including two Dirac operators:
\begin{align}
\{\{\DGO,a\},\{\{\DGO,b\},c\}\}
=&-(-1)^{|a|}\{\{ \DGO, \{\{\DGO,a\},b\}\},c\}+(-1)^{(|a| +1)(|b| +1)}\{\{\DGO,b\},\{\{\DGO,a\},c\}\}
\cr~~~~&~~~~~~~~~~~~~~~~~~~~~~
+(-1)^{|a| }\frac{1}{2}\{\{\{\{\DGO,\DGO\},a\},b\},c\}.
\label{Jacobiofderivedbracket01}
\end{align}
By using (\ref{pDerivedBracketforCA1}), the above identity gives the following relation for $a,b,c\in\Gamma(E)$:
\be
[a,[b,c]]
=[[a,b],c]+[b,[a,c]]-\frac{1}{2}\{\{\{\{\DGO,\DGO\},a\},b\},c\}.
\label{Jacobiofderivedbracket01}
\ee
If the last term is zero, the above relation gives the graded Jacobi identity of the derived bracket.
Compared to the formulation of DFT using the graded manifold approach \cite{Carow_Watamura_2019}, this part corresponds to the weak master equation.

Since we do not require the Jacobi identity in the metric algebroid, the last term 
gives a measure for the breaking of the Jacobi identity. We obtain the following representation of 
$\GL$ and $\GL'$ defined in (\ref{pCAJacobiator}) and (\ref{pCAJacobiator2}), respectively,
 by the derived bracket:
\benu
\item $\GL$: for $a,b,c\in\Gamma(E)$
\be
\GL(a,b,c)=-\{\{\{\DGO^2,a\},b\},c\}.\label{WeakmasterDGO1}
\ee
\item $\GL'$: for $a,b\in\Gamma(E)$ and $f\in C^\infty(\mathbb{M})$ 
\be
\GL'(a,b)f=\{\{\{\DGO^2,a\},b\},f\}.\label{WeakmasterDGO2}
\ee
\eenu

\subsubsection{Generalized Lie derivative}\label{pGeneralizedLieDerivativesection}
Since in the above definition, the bracket $[e , a]$ for $e,a\in \Gamma(E)$ is defined by the graded commutator of 
$\{\DGO,e\}$ and $a$ as in (\ref{pDerivedBracketforCA1}), it is natural to extend the bracket 
to any element $a\in \Gamma(Cl(E))$ of the Clifford bundle, and define a derivative $\CL_e$ on $\Gamma(Cl(E))$ 
by
\be
\CL_e  a=\{\{\DGO,e\}, a\}~.
\ee
We call $\mathcal{L}_e$ a generalized Lie derivative.

The action of the generalized Lie derivative can be extended to the spin bundle by requiring the 
Leibniz rule: for $a\in\Gamma(Cl(E))$ and $\chi\in\Gamma(\mathbb{S})$ 
\be
\mathcal{L}_e a\chi=(\mathcal{L}_e a)\chi+ a\mathcal{L}_e\chi~.
\ee
The generalized Lie derivative on $\chi$ apparently satisfies the above Leibniz rule w.r.t. the Clifford action:
\be
\mathcal{L}_e \chi=\{\DGO,e\}\chi~.
\label{GLonS}
\ee 
Note that there is an ambiguity to add a function.
In particular, the closure of the Lie derivative on spinor yields
\bea
\{\mathcal{L}_a,\mathcal{L}_b\}\chi
&=&\{\{\DGO,a\},\{\DGO,b\}\}\chi
\cr&=&
\{\DGO,\{\{\DGO,a\},b\}\}\chi
-\{\{\DGO^2,a\},b\}\chi
\cr&=&
\{\DGO,[a,b]\}\chi
-\{\{\DGO^2,a\},b\}\chi
\cr&=&
\mathcal{L}_{[a,b]}\chi
-\{\{\DGO^2,a\},b\}\chi~.
\eea

\subsection{Dirac generating operator}
\label{section_DGO}
In this section, we give a concrete form of the Dirac generating operator using a local basis.
Namely, we construct a Dirac operator which satisfies the conditions 
(\ref{pDerivationforCA1}),(\ref{pDerivedBracketforCA1}) and (\ref{pAnchorforCA1}).

We use the standard representation of the Clifford action defined by
\be
\{\gamma_A,\gamma_B\}=2\eta_{AB}~.\label{CliffordCommu}
\ee
We also use a zero connection $\partial_A$: $\Gamma(\mathbb{S})\rightarrow\Gamma(\mathbb{S})$ defined by
\bea
\{\partial_A,f\}&=&\rho(E_A)f~,
\\\{\partial_A,\gamma_B\}&=&0~,\label{pDiffgamma}
\\\partial_A\Ket{0}&=&0~,
\eea
where $\Ket{0}\in \mathbb{S}$ is a pure spinor, see appendix (\ref{def_purespinor}), and $f\in C^\infty$.
Since the metric $\eta_{BC}$ is constant, (\ref{pDiffgamma}) is compatible with (\ref{CliffordCommu}).

A general form of the Dirac generating operator is given by the following odd differential operator
\be
\DGO=\half(\gamma^A\p_A-{1\over12}F_{ABC}\gamma^{ABC}- \half F_A\gamma^A)~,
\label{pGeneralformofDGOeq}
\ee
where $\gamma^{ABC}={1\over6}\gamma^{[A}\gamma^{B}\gamma^{C]}$. 
Then,
it is straightforwards to show that $\DGO$ satisfies conditions 
(\ref{pDerivationforCA1}),(\ref{pDerivedBracketforCA1}) and (\ref{pAnchorforCA1}).
The structure function $F_A$ is an ambiguity of the Dirac generating operator, i.e., the metric algebroid is independent of the choice of the structure function $F_A$.
\footnote{Note that if we have another Dirac generating operator $\tilde{\DGO}$ which satisfies
the conditions (\ref{pDerivationforCA1}),(\ref{pDerivedBracketforCA1}) and (\ref{pAnchorforCA1}),
the difference $\slashed D-\tilde{\slashed D}$ satisfies
\bea
&&\quad \{\{\slashed D-\tilde{\slashed D},f\}=0~~,\\
&&\quad \{\{\slashed D-\tilde{\slashed D},a\},b\}=0~~.
\eea
Thus, since $\DGO$ is a odd graded operator, we are free to choose $F_A\in C^\infty$,
 i.e., $\slashed D-\tilde{\slashed D}\in\Gamma(E)$.}

A representation of the connection $\nabla^{\mathbb{S}}$ on $\chi\in\Gamma(\mathbb{S})$ is specified by the action on the pure spinor $\Ket{0}$. 
The relations (\ref{pdef:inducedconnectiononS}) and  (\ref{pdef:inducedconnectiononSf}),
are realized by taking 
\be
\nabla^{\mathbb{S}}_A\Ket{0}=(-{1\over4}W_{ABC}\gamma^{BC}+\half A_A)\Ket{0}~,
\ee
or equivalently, by defining the connection as
\be
\nabla^{\mathbb{S}}_A=\p_A-{1\over4}W_{ABC}\gamma^{BC}+\half A_A~.
\ee
The Dirac operator with this connection is
\bea
\gamma^A\nabla_A^{\mathbb{S}}
&=&\gamma^A\p_A-
{1\over24}W_{[ABC]}\gamma^{ABC}
+(\half A_A-\half W_B{}^B{}_A)
\gamma^A~.
\label{SimplestDiraconS}
\eea
Therefore, using the connection $\nabla^{\mathbb{S}}$ we can write the Dirac generating operator as
\be
\DGO=\half\gamma^A\nabla_A^\mathbb{S}
+{1\over24}(\half W_{[ABC]}-F_{ABC})\gamma^{ABC}-{1\over4}(F_A+A_A- W^B{}_{BA})\gamma^A~.
\ee
From the metric algebroid point view, the last term which is proportional to $\gamma^A$
 is an ambiguity. We use it in such a way that $F_A$ coincides with the trace of the connection,
 i.e., $F_A=W^B{}_{BA}$ and $A_A=0$, which is convenient from the point of view of DFT.

Using the definition of the $E$-torsion, $\DGO$ can be written as
\be
\DGO=\half\gamma^A\nabla_A^\mathbb{S}
+{1\over24}T_{ABC}\gamma^{ABC}~.\label{pDGOcovtorsionfree}
\ee
This form shows that the Dirac generating operator is the Dirac operator with
torsion free connection $W'_{ABC}$ in (\ref{TorsionCancelledConnection}).
As in \cite{Garcia-Fernandez:2016aa}, the Dirac generating operator is characterized by the structure functions $F_{ABC}$ and $F_A$,
thus the $E$-connection in (\ref{pDGOcovtorsionfree}) is not determined uniquely.

In the standard DFT, the Dirac operator in the same form as (\ref{pGeneralformofDGOeq}) is used in \cite{Geissbuhler:2013uka} where the structure functions 
$F_{ABC}$ and $F_A$ are replaced by $\CF_{ABC}$ and $\CF_A$ to formulate the action of the Ramond-Ramond sector \cite{Hohm:2011aa, Hohm_2011, Jeon:2012aa}. It is also used to formulate the Ramond-Ramond sector of 
DFT on Drinfeld double \cite{Hassler:2017yza}.

\subsection{Generalized Lichnerowicz formula and pre-Bianchi identity}

From the metric algebroid point view, DFT belongs to a class which is specified by the
pre-Bianchi identity. We show that the conditions corresponding to the pre-Bianchi identity can be 
derived by using a generalized Lichnerowicz formula.
The Lichnerowicz formula is a relation formulated by the difference of the square of the Dirac operator 
and a Laplace operator, cancelling the differential operators.
Here, we define the generalized Lichnerowicz formula for the metric algebroid 
using the Dirac generating operator given in (\ref{pDGOcovtorsionfree})
which is induced by the $E$-connection but torsion free.

\subsubsection{Divergence on spin bundle and Laplace operator}

In order to define the Laplace operator, first we introduce here the divergence operator on 
the spin bundle.
We define a contraction  $\iota_{e_1}: \Gamma(E)\otimes \Gamma(\mathbb{S})\rightarrow\Gamma(\mathbb{S}) $
 of $e_1\in\Gamma(E)$ with $e_2\otimes \chi\in \Gamma(E)\otimes\Gamma(\mathbb{S})$ as
\be
\iota_{e_1}(e_2\otimes\chi)=\braket{e_1,e_2}\chi~.
\ee
The connection can be considered as a map 
$\nabla^\mathbb{S} : \Gamma(\mathbb{S})\rightarrow \Gamma(E)\otimes\Gamma(\mathbb{S})$, then
\be
\iota_e\nabla^\mathbb{S}\chi=\nabla_e^\mathbb{S}\chi~.
\ee
An associated connection on the tensor product $\nabla^{E\otimes\mathbb{S}}:\Gamma(E)\times\Gamma(E)\otimes\Gamma(\mathbb{S})\rightarrow\Gamma(E)\otimes\Gamma(\mathbb{S})$ is given
by the Leibniz rule
\be
\nabla_{e_1}^{E\otimes S}{e_2}\otimes \chi=(\nabla^E_{e_1}e_2)\otimes \chi+e_2\otimes \nabla^{\mathbb{S}}_{e_1}\chi,
\ee
where $e_1,e_2\in \Gamma(E), \chi\in \Gamma(\mathbb{S})$.

The divergence on the spin bundle is then defined by applying the definition given in \cite{AlekseevDGO,Garcia-Fernandez:2016aa,Severa:2018aa} as a map 
$div:\Gamma(E)\otimes\Gamma(\mathbb{S})\rightarrow \Gamma(\mathbb{S})$ 
satisfying the following relation:
\be
div(fe\otimes\chi)
=
(\rho(e)f)\chi+fdiv(e\otimes\chi)~,\label{pAxiomDivergenceonSpinRep}
\ee
for any $f\in C^\infty(\mathbb{M})$.

For a given $E$-connection, we can define a divergence $div_\nabla$ by using a local basis as
\bea
div_\nabla(e\otimes\chi)
&=&\iota_{E^A}\nabla^{E\otimes\mathbb{S}}_{E_A}(e\otimes\chi)
\cr&=&\braket{{E^A},\nabla_{E_A}^Ee}\chi+\braket{E^A,e}\nabla^\mathbb{S}_{E_A}\chi~.
\eea
It is clear that this satisfies the above condition (\ref{pAxiomDivergenceonSpinRep}).
In the appendix we show that the following $div^U_\nabla$ 
 also satisfies the above condition of a divergence (\ref{pAxiomDivergenceonSpinRep}):
\bea
div^U_\nabla=div_\nabla(e\otimes\chi)-\iota_{U}(e\otimes \chi)~,
\eea
where $U\in\Gamma(E)$, showing the degree of freedom in the divergence.
The Laplacian of the given $E$-connection $\nabla^E$ is defined by
\be
\Delta\chi
=div^U_\nabla\nabla\chi~.
\ee
Since the divergence has an ambiguity, the Laplacian has also an ambiguity of $U\in\Gamma(E)$.

\subsubsection{Generalized Lichnerowicz formula}\label{pSectiongeneralizedLichnero}

The Laplace operator which appears in the generalized Lichnerowicz formula is
the one associated to the connection $\nabla^{\phi'}$ defined
by $\phi'(a,b,c)$ in (\ref{pDefofphiprime}). 
We have shown that the structure function $\phi'_{ABC}$ has the same transformation 
property as the connection $W_{CAB}$, thus 
we introduce a connection on the Clifford module $\Gamma(\mathbb{S})$ s.t.
\be
\nabla^{\phi'}_A=\p_A-{1\over4}\phi'_{BCA}\gamma^{BC}~.
\ee
The corresponding Laplace operator is then given by
\be
\Delta^{\phi'}=div^U_{\nabla^{\phi'}}\nabla^{\phi'}=\eta^{AB}\nabla_A^{\phi'}\nabla_B^{\phi'}-(\phi'_{B}{}^{AB}+U^A)\nabla^{\phi'}_A~,
\ee
where $U_A$ is a vector representing the ambiguity in the divergence as discussed previously. 
The definition of the generalized Lichnerowicz formula 
is the square of the Dirac generating operator with derivative terms covariantly subtracted:
\begin{align}
4\DGO^2-\Delta^{\phi'}=&
-{1\over24}F_{ABC}F^{ABC}-\half(\rho(E^A)F_A)+(-F^A+\phi'_{E}{}^{AE}+U^A)\p_A
+{1\over4}F_AF^A
+{1\over8}\phi'_{BCA}\phi'^{BCA}
\cr&
+{1\over 4}\left(-\CJ_{BCD}{}^D
+ (\rho(E_{[B})(-F_{C]}+\phi'^D{}_{C]D})-(-F^A+\phi'_D{}^{AD})F_{ABC}-U_A\phi'_{BC}{}^A
\right)\gamma^{BC}
\cr&
-{1\over2}\tilde\phi_{BCB'C'}\gamma^{BCB'C'}
\cr=&
{1\over4}R^\nabla_{AB}{}^{AB}
+(-W_B{}^{BA}+\phi'_{B}{}^{AB}+U^A)\p_A
+{1\over8}\phi'_{BCA}\phi'^{BCA}
\cr&
-{1\over 4}\Big(
\CJ_{BCD}{}^D
+ (\rho(E_{[B})(W^B{}_{B|C]}-\phi'^D{}_{C]D})-(W_B{}^{BA}-\phi'_D{}^{AD})F_{ABC}+U_A\phi'_{BC}{}^A
\Big)\gamma^{BC}
\cr&
-{1\over48}\tilde\phi_{BCB'C'}\gamma^{BCB'C'}~,\label{pGeneralizedLichnerowicz1}
\end{align}
where $\CJ_{BCD}{}^D$ is the tensor defined in the identity given in (\ref{pJacobionTM}).
The term proportional to $\gamma^{ABCD}$ is the tensor $\tilde\phi_{ABCD}$ in (\ref{pDefphitilde}) 
which gives the pre-Bianchi identity.

The derivative terms of the square of the Dirac operator appear in both, the scalar part and 
 the part proportional to $\gamma^{AB}$.
The Laplacian $\Delta^{\phi'}$ is chosen such that the terms containing the derivative operator in 
the part $\gamma^{AB}$ cancel.  However, the derivative term in the scalar part 
remains as shown in the first line on the r.h.s.
Now, the freedom $U_A$ in the divergence is used to compensate this derivative term including the trace of the connection
$W^B{}_{BA}=F_A$. This means that we obtain the relation $F_A=\phi'_{BA}{}^B+U_A$ so that the derivative terms in the scalar part on the r.h.s. vanish.
After this identification of $U_A$ in terms of the flux $F_A$, the second line on the r.h.s. of (\ref{pGeneralizedLichnerowicz1})
becomes $\CJ_{BCD}{}^D$ plus the l.h.s. of (\ref{pGSEgeneralizedvector}) which gives
the pre-Bianchi identity for the flux $F_A$ in (\ref{pBianchiofOneIndex1}).
As discussed in (\ref{pSingleStructurefunctionindilaton}), $U_A=2\rho(E_A)d$ satisfies the pre-Bianchi
identity.

Finally the scalar part is given by the generalized Riemann scalar $\BR$ constructed from the 
generalized curvature $\CR_{ABCD}=\CR(E_A,E_B,E_C,E_D)$ in (\ref{pCurvatureinprepreCA}):
\be
\BR=\CR_{AB}{}^{AB}=2R^\nabla_{AB}{}^{AB}+\phi'_{ABC}\phi'^{ABC}~.\label{pCovariantcurvaturescalar1}
\ee
Recall that the Dirac generating operator (\ref{pGeneralformofDGOeq}) is
defined by the fluxes $F_{ABC}$ and $F_A$, which do not define the connection uniquely. 
Furthermore, 
the generalized curvature tensor $\CR_{ABCD}$ in (\ref{generalizedcurvatureinlocalbasis})
 is expressed by the $E$-connection $W_{ABC}$ which is not completely determined by the flux.
 However, from (\ref{pDGOcovtorsionfree}) we know that the connection in the Dirac generating operator is the torsion free connection $W'_{ABC}$ in (\ref{TorsionCancelledConnection}), 
and $F_{ABC}=\half W'_{[ABC]}$. 
Therefore, $R^\nabla$ in the generalized Lichnerowicz formula which is written with the connection $W'$, is
represented by the structure functions as    
\bea
R^\nabla_{AB}{}^{AB}(W')
=-{1\over 6}F_{ABC}F^{ABC}
-2\rho(E_{A})F^A
+F^AF_A~,
\label{relationFFandRfortorsion0}
\eea
where the ambiguity of the Dirac generating operator is identified as $F_A=W'^B{}_{BA}=W^B{}_{BA}$

As we discussed, DFT is realized on a metric algebroid where the pre-Bianchi identities
vanish. 
For this class of metric algebroid, we have the following 
 generalized Lichnerowicz formula:
\be
4\DGO^2-\Delta^{\phi'}={1\over8}\BR~.\label{pGeneralizedLichnerowiczformula1}
\ee

The above result can be put into the following statement:
The requirement that the pre-Bianchi identities for the structure functions hold
 can be rephrased as the requirement that the generalized Lichnerowicz formula is satisfied.
Note that the generalized scalar curvature does not vanish in general.

\section{Action from Dirac generating operator}

To construct an action using the above approach, we propose a projected generalized Lichnerowicz formula which is consistent with Riemannian structure on the metric algebroid. 

\subsection{Riemannian structure}
The splitting of the vector bundle in DFT and in generalized geometry has been 
worked out in great detail, as can be found in  \cite{Jeon:2010rw,Jeon:2011aa,Hohm:2011ab,Hohm_2012,Hohm_2013,Coimbra_2011}.

It is known that the metric structure on DFT can be introduced by splitting the 
vector bundle $E$ into positive and negative sub-bundle $V^\pm$ as in the generalized geometry:
\be
E=V^+\oplus V^-~,
\ee
with
\be
V^+=\{a\in E~|~\braket{a,a}=|\braket{a,a}|\}
~~,~~V^-=\{a \in E~|~\braket{a,a}=-|\braket{a,a}|\}~,
\ee 
where $V^+$ and $V^-$ are orthogonal to each other.

Using the projection operators $\CP^\pm : E\rightarrow V^\pm$, any vector can be
split into $V^\pm$ as $a=a^++a^-$ where $a^\pm=\CP^\pm(a)$. The sub-bundles $V^+$ and
$V^-$ are orthogonal and thus  
the inner product can be split as
\be
\braket{a,b}=\braket{a,\CP^+(b)}+\braket{a,\CP^-(b)}=\braket{a^+,b^+}+\braket{a^-,b^-}~.
\ee  
The generalized metric is a positive definite product defined 
for $a,b \in E$ by
\be
\CH(a,b)=\braket{a^+,b^+}-\braket{a^-,b^-}~.\label{pGeneralizedMetricasmap}
\ee
By the identification of the dual space $E^*$ with $E$ via the metric $\braket{-,-}$,
$\CH$ can be considered as a map $\CH=\CP^+-\CP^-:E\rightarrow E$ (see appendix \ref{compatibledivergenceappend}). 
By using a local basis we have
\be
\CH(E_A)=\CH_{A}{}^BE_B~,
\ee 
where $\CH_{AB}=\CH_{BA}$ and $\CH_A{}^B\CH_B{}^C=\delta_A{}^C$. Then, we can define 
projection operators $\CP^{\pm}$ to the
positive/negative sub-bundle w.r.t. the generalized metric $\CH$ by
\be
\CP^\pm=\half(\id\pm \CH)~,\label{pDefprojectionpm} 
\ee 
By compatibility with the metric structure, the structure group $O(D,D)$ reduces to $O(D)\times O(D)$.
A local basis is chosen such that the projection operators become diagonal,
denoted by $E_A=(E_{a},E_{\bar a})\in V^-\oplus V^+$ with 
\be
\CP^+(E_{ a})=E_{ a}~~,~~\CP^-(E_{\bar a})=E_{\bar a}~,
\ee 
and
\be
\braket{E_a,E_b}=\Gs_{ab}~,~~\braket{E_{\bar a},E_{\bar b}}=-\bar \Gs_{\bar a\bar b}~,~~\braket{E_a,E_{\bar b}}=0~.
\ee
The above relation defines the explicit form of the $O(D,D)$ metric as
\bea
\braket{E_{ A},E_{ B}}
&=&\eta_{ A B}
=\Matrix{cc}{\eta_{ab}&0\cr0& \eta_{\bar a\bar b}}
=\Matrix{cc}{\Gs_{ab}&0\cr0& -\bar\Gs_{\bar a\bar b}}~,
\eea
where $\Gs_{ab}=\bar\Gs_{\bar a\bar b}$ are local Lorentz metric.  The corresponding basis of the Clifford bundle
 is also split as $\gamma_A=(\gamma_a,\gamma_{\bar a})$, and their commutation relation is given by (\ref{CliffordCommu}).
 
\subsection{Compatible connection}
As in the generalized geometry, we consider the $E$-connection $\nabla^E: \Gamma(E)\times \Gamma(V^\pm)\rightarrow\Gamma(V^\pm)$ compatible with the splitting.
Using the local basis $E_A\in\Gamma(E)$, 
compatibility requires
\be
\braket{\nabla^E_{E_A}E_b,E_{\bar c}}=0~~,~
\braket{\nabla^E_{E_A}E_{\bar b},E_{ c}}=0~.
\ee
From this we conclude that the nonzero components of the connection are $W_{Aab}$ and $W_{A\bar a\bar b}$. 

In the following, we construct an action from the Dirac generating operator.
As we discussed, the Dirac generating operator is free from torsion as given in (\ref{pDGOcovtorsionfree}) and thus we can choose the torsionless connection 
without loosing generality\footnote{Torsion terms can be recovered replacing the torsionless connection by $W'_{ABC}
=W_{ABC}-{1\over3}T_{ABC}$.}. Then we get the relation between the structure function and connection as
\be
F_{ABC}=\half W_{[ABC]}~,
\ee
i.e., the totally antisymmetric part of the connection is defined by the structure function.

From the definition of the torsion (\ref{pEtorsionPreCA}), for the mixed argument we get
\bea
T(a^-,b^+,c^+)
&=&\braket{\nabla^E_{a^-}b^+-\nabla^E_{b^+}a^--[a^-,b^+], c^+}+\braket{\nabla^E_{c^+}a^-,b^+}
\cr &=&\braket{\nabla^E_{a^-}b^+-[a^-,b^+], c^+}~.
\eea
and similarly for $T(a^+,b^-,c^-)$.
Thus the mixed part of the torsionless compatible connection is given by
\be
\nabla_{a^-}^Eb^+=\CP^+[a^-,b^+]~,
\ee
which is known as the generalized Bismut connection in generalized geometry \cite{Gualtieri:aa,Hitchin_2011}.
From this, we conclude that the mixed part of the compatible connection is completely defined by the structure
function as
\be
W_{\bar abc}=F_{\bar abc}~~,~~W_{a\bar b\bar c}=F_{a\bar b\bar c}~.
\ee
On the other hand, the pure part of the spin connection is not completely defined by the structure functions 
in the Dirac generating operator 
except for the totally antisymmetric components: 
\be
\half W_{[abc]}=F_{abc}~~,~~\half W_{[\bar a\bar b\bar c]}=F_{\bar a\bar b\bar c}~,
\ee
and the trace part 
\be
W^B{}_{BA}=F_A~,
\ee
and thus
\be
W^B{}_{Ba}=F_a~~,~~W^B{}_{B\bar a}=F_{\bar a}~.
\ee

\subsection{Projected Dirac operator and Laplacian}

In the following we formulate the action using the generalized Lichnerowicz formula
with the above compatible connection. For this we introduce here the projected Dirac operator and the Laplacian.

First, we consider the connection $\nabla^{\mathbb{S}^+}:\Gamma(\mathbb{S}^+)\rightarrow\Gamma(E^*)\otimes\Gamma(\mathbb{S}^+)$, which is defined as
\bea
\nabla_A^{\mathbb{S}^+}=\p_A-{1\over4}W_{Abc}\gamma^{bc}~,
\eea
where $\Gamma(\mathbb{S}^+)$ is a module over $Cl(V^+)$, constructed on $\Ket{0}$ by multiplying the elements of $\Gamma(V^+)$.
(See appendix for details.)
Then, we consider the following projected connections
\bea
\nabla_+^{\mathbb{S}^+}&=&E^a\otimes \nabla_a^{\mathbb{S}^+}~,
\cr\nabla_-^{\mathbb{S}^+}&=&E^{\bar a}\otimes \nabla_{\bar a}^{\mathbb{S}^+}~.
\label{pProjectedConnectiononspin+-}
\eea

These connections are invariant under $O(D)\times O(D)$ rotation of the basis of $E$ 
and covariant w.r.t. the local $O(D)$ rotation of $\mathbb{S}^+$.
The corresponding Dirac operator on $\Gamma(\mathbb{S}^+)$
can be written by the structure functions $F_{abc}$ and $F_a$ as
\bea
\DGO^+&=&\half \gamma^a\nabla_a^{\mathbb{S}^+}
\cr&=&\half \gamma^a\p_a-{1\over24}F_{abc}\gamma^{abc}-{1\over4} F_a\gamma^a~.
\label{pProjectedDGO}
\eea
Since $\nabla^{\mathbb{S}^+}_{\bar a}$ contains the only mixed type $E$-connection, it can also be written by the structure function as
\be
\nabla^{\mathbb{S}^+}_{\bar a}=\p_{\bar a}-{1\over 4}\gamma^{bc}F_{\bar abc}~.
\label{pProjectedconnection+}
\ee

As in the generalized Lichnerowicz formula, we further have to consider the connection induced by $\phi'_{ABC}$ on the Clifford module $\Gamma(\mathbb{S}^+)$ s.t.
\be
\nabla^{\phi'^+}_A=\p_A-{1\over4}\phi'_{bcA}\gamma^{bc}~.
\ee
The corresponding Laplace operator is given as
\be
\Delta^{\phi'^+}=div^U_{\nabla^{\phi'}}\nabla^{\phi'^+}=\eta^{AB}\nabla^{\phi'^+}_{A}\nabla^{\phi'^+}_{B}-(\phi'_{B}{}^{AB}+U^A)\nabla^{\phi'^+}_A~.\label{Laplacianphiprime+}
\ee

\subsection{Projected Lichnerowicz formula and DFT action}

The action of DFT can be formulated by the following projected Lichnerowicz formula as
\bea
L=4\DGO^{+2}+div_\nabla \nabla_-^{\mathbb{S}^+}-\Delta^{\phi'^+}~.
\eea
The first two terms are the analogous combination appearing in \cite{Coimbra_2011} where the supergravity is formulated using the generalized geometry. 
The difference is that the first two terms here contain the differential operators.

By using the projected Dirac operator $\DGO^+$ in (\ref{pProjectedDGO}), the first term is
\bea
4\DGO^{+2}&=&\partial^a\partial_a-F^a\partial_a-\frac{1}{24}F_{abc}F^{abc}-\frac{1}{2}\rho(E_a)F^a+\frac{1}{4}F_aF^a
\cr&&+\gamma^{ab}(\frac{1}{2}\partial_{[a}\partial_{b]}-\frac{1}{2}F_{ab}{}^c\partial_c-\frac{1}{4}\partial_cF_{ab}{}^c+\frac{1}{4}F^cF_{abc}-\frac{1}{4}\rho(E_{[a})F_{b]})
\cr&&+\gamma^{abcd}(\frac{1}{384}F_{e[ab}F^e{}_{cd]}-\frac{1}{288}\rho(E_{[a})F_{bcd]})~.
\eea
The second term is the divergence of the projected connection 
$\nabla_-^{\mathbb{S}^+}$ given in (\ref{pProjectedConnectiononspin+-}):
\bea
div_\nabla \nabla_-^{\mathbb{S}^+}
&=&div_\nabla(E^{\bar a}\otimes \nabla_{\bar a}^{\mathbb{S}^+})
\cr&=&
\partial_{\bar a}\partial^{\bar a}-F^{\bar a}\partial_{\bar a}-\frac{1}{8}F^{\bar{a}bc}F_{\bar{a}bc}
\cr&&+\gamma^{ab}(-\frac{1}{2}F_{ab}{}^{\bar c}\partial_{\bar c}+\frac{1}{4}F^{\bar c}F_{ab\bar c}-\frac{1}{4}\rho(E_{\bar c})F_{ab}{}^{\bar c})
\cr&&+\frac{1}{384}\gamma^{abcd}F_{[ab}{}^{\bar e}F_{cd]\bar e}~,
\eea
where $div_\nabla$ of the projected connection is given in the appendix.
The last term is the Laplacian from the projected connection $\nabla^{\phi'^+}$ given in (\ref{Laplacianphiprime+}):
\bea
\Delta^{\phi'^+}&=&\partial_A\partial^A-(\phi'_B{}^{AB}+U^A)\partial_A-\frac{1}{8}\phi'_{abC}\phi'^{abC}
\cr&&+\gamma^{ab}(-\frac{1}{2}\phi'_{ab}{}^C\partial_C-\frac{1}{4}\rho(E_C)\phi'_{ab}{}^C+\frac{1}{4}
(\phi'_{BC}{}^B+U_C)\phi'_{ab}{}^C)
\cr&&+\frac{1}{384}\gamma^{abcd}\phi'_{[ab}{}^E\phi'_{cd]E}~.
\eea

As in the generalized Lichnerowicz formula derived in \S\ref{pSectiongeneralizedLichnero}, the last term cancels the differential operators in the first two terms keeping the covariance.
For this we identify the vector field $U_A$,i.e., the ambiguity in the divergence, as in the case of 
the generalized Lichnerowicz formula as
\be
F_A=\phi'_{BA}{}^B+U_A~.
\ee
The result is 
\bea
L=\CR^{DFT}-{1\over4}\Big(\CJ_{abC}{}^C+ \rho(E_{[a})(U_{b]})-U^C(F_{Cab}-\phi'_{abC})\Big)\gamma^{ab}-\half\tilde\phi_{abcd}\gamma^{abcd}~,
\eea
where 
\bea
\CR^{DFT}=-\frac{1}{24}F_{abc}F^{abc}-\frac{1}{8}F^{\bar{a}bc}F_{\bar{a}bc}-\frac{1}{2}\rho(E_a)F^a+\frac{1}{4}F_aF^a-\frac{1}{8}\phi'_{abC}\phi'^{abC}~.
\eea

By using the identification of the structure functions $F_{abc}$, $F_A$ and spin connection $W_{Abc}$, we can write the above $\CR^{DFT}$
as
\be
\CR^{DFT}=\CR_{ABCD}\CP^{+AC}\CP^{+BD}~,
\ee
where the $\CR_{ABCD}$ is the generalized curvature in the metric algebroid defined in (\ref{generalizedcurvatureinlocalbasis}).

The above action is $O(D)\times O(D)$ covariant. 
Up to the section condition, we can see
 that the $\CR^{DFT}$ is proportional to the standard DFT action
  if we substitute 
$F_{ABC}=\CF_{ABC}$ and $\phi'_{ABC}=\Omega_{CAB}$,
which is a solution of the pre-Bianchi identity (\ref{explicitformofpreBianchi}). 
We also identify the dilaton by  (\ref{pSingleStructurefunctionindilaton}), i.e., 
the identification $F_A=\CF_A$ where
 \be
 \CF_A= \Omega^B{}_{BA}+2\rho(E_A)d~,\label{pStandardDFTfluxwithsingleflux}
 \ee
which is also a solution of the pre-Bianchi identity (\ref{pBianchiofOneIndex1}).

Now, we can formulate the action of the DFT using the above projected Lichnerowicz formula.
First, we introduce an inner product $(-,-)_{\mathbb{S}}:\Gamma(\mathbb{S})\times\Gamma(\mathbb{S})\rightarrow\Gamma(\Lambda)$ as discussed in appendix \ref{AppendixDilatondensity},
for $f\in C^\infty(\mathbb{M})$, $\chi_1,\chi_2\in\mathbb{S}$ and $a\in \Gamma(E)$,
\bea
(f\chi_1,\chi_2)_{\mathbb{S}}&=&(\chi_1,f\chi_2)_\mathbb{S}=f(\chi_1,\chi_2)_\mathbb{S},
\\(\chi_1,a\chi_2)_\mathbb{S}&=&(a\chi_1,\chi_2)_\mathbb{S},
\\\CL_a(\chi_1,\chi_2)_\mathbb{S}&=&(\CL_a\chi_1,\chi_2)_\mathbb{S}+(\chi_1,\CL_a\chi_2)_\mathbb{S}~,
\eea
where $\Lambda=(\Lambda^\frac{1}{2})^2$. 
Note that the action of the generalized Lie derivative on 
this inner product is equal to the one for $\Gamma(\Lambda)$.
From this definition $\braket{0|0}:=(\Ket{0},\Ket{0})_\mathbb{S}$ is invariant 
under the $O(D,D)$ rotation.
Thus, we obtain an $O(D)\times O(D)$ invariant combination $S_{inv}$,
\bea
S_{inv}&:=&\Bra{0}(4\DGO^{+2}+div_\nabla \nabla_-^{\mathbb{S}^+}-\Delta^{\phi'^+})\Ket{0}
\cr&=&\braket{0|0}\CR^{DFT}
\cr&=&C\mu_0 e^{-2d}\CR^{DFT}~,
\eea
where $C=\braket{0'|0'}$ is a constant and $\mu_0$ is defined in appendix \ref{AppendixDilatondensity}.

Applying this formulation to the standard DFT, $S_{inv}$ becomes equivalent to the standard action
as follows.
First, we construct a concrete representation of $\Gamma(\Lambda^{\frac{1}{2}})$ where $\Lambda^{1\over2}$ is 
a line bundle defined in (\ref{LinebundleLambda}), and 
$\Gamma(\Lambda^{\frac{1}{2}})$ is characterized by the action of the generalized Lie derivative on $f\mu^\frac{1}{2}\in\Gamma(\Lambda^{\frac{1}{2}})$
\be
\CL_a(f\mu^\frac{1}{2})=(\rho(a)f)\mu+f(-\rho(a)d+\frac{1}{2}(\partial_Na^N))\mu^\frac{1}{2}.
\ee
If this equation is satisfied, the choice of the representation of $\Gamma(\Lambda^{\frac{1}{2}})$ is not relevant for the algebroid structure.
Then, we can identify $\Gamma(\Lambda^{\frac{1}{2}})$ with $\Gamma\Big((\wedge^{top}T\mathbb{M})^\frac{1}{2}\Big)$ as follows:
\bea
\mu^\frac{1}{2}&=&e^{-d}(dx^1\wedge\cdots\wedge dx^{2D})^\frac{1}{2}~,
\\\CL_a(f\mu^\frac{1}{2})&=&L_a(f\mu^\frac{1}{2})~,
\eea
where $L_a$ is the standard Lie derivative on $\Gamma(\Lambda^{\frac{1}{2}})$ with $\Gamma\Big((\wedge^{top}T\mathbb{M})^\frac{1}{2}\Big)$.
Thus, the generalized Lie derivative on $\Gamma(\Lambda^{\frac{1}{2}})$ is defined by
the standard Lie derivative in this representation.
Using this representation, $S_{inv}$ becomes
\bea
S_{inv}&=&\mu\braket{0'|0'}\CR^{DFT}
\cr&=&C\int dx^1\wedge\cdots\wedge dx^{2D}e^{-2d}\CR^{DFT}.
\eea
where $\mu_0$ is identified with the integration over $\mathbb{M}$.
We can prove that $S_{inv}$ is invariant under the generalized Lie derivative as follows,
\be
\CL_aS_{inv}=C\int dx^1\wedge\cdots\wedge dx^{2D}\partial_N(a^Ne^{-2d}\CR^{DFT})=0~.
\ee
This action is equal to the standard action under the section condition.

On the other hand, the combination 
\be
\tilde{S}_{int}=C\mu_0fe^{-2d}\CR^{DFT},\ (f\in C^\infty(\mathbb{M}))
\ee
is also invariant under the $O(D)\times O(D)$ rotation.
To fix the ambiguity $f$, we need to discuss
the gauge transformation of the field. 
In particular, since the dilaton $d$ is considered as a function in $C^\infty(\mathbb{M})$ here, 
there is a difference between the generalized Lie derivative and the gauge transformation. 
As discussed in the appendix 
\ref{AppendixDilatondensity},
the action of the generalized Lie derivative $\CL_a$ of the present formulation 
generates the gauge transformation $\hat{\delta}_a$ for the dilaton as 
\be
\CL_ae^{-d}\mu_0^\frac{1}{2}=(\hat{\delta}_ae^{-d})\mu_0^\frac{1}{2}~.
\ee
The transformation $\hat\delta_a$ can be identified with the gauge transformation of the
 dilaton in the standard DFT and in this way $e^{-d}$ is considered as a half density. 
The gauge transformation of the field can be discussed by choosing a concrete form of 
the structure functions $F_{ABC}$ and $ F_A$, and discussing the failure of the covariance.
For example, by taking the standard DFT solution of the pre-Bianchi identity,
i.e. $F_{ABC}=\CF_{ABC}$ and $F_A=\CF_A$, it is known that the action is gauge invariant (see for example \cite{Aldazabal_2013}).

\section{Closure and derived bracket}
In the formulation of a Courant algebroid using the Dirac generating operator, the closure of the bracket, i.e. the Jacobi identity, is realized by requiring that the square of the Dirac generating operator is a function \cite{AlekseevDGO}.  
Here, we are considering a metric algebroid, i.e.,
the Jacobi identity is not required for the derived bracket and thus
the square of the Dirac operator is not necessarily a function. 
On the other hand, since the gauge symmetry of DFT is generated by the generalized Lie derivative
\cite{Hull:2009aa},
the closure of the D-bracket on the fields, which is the gauge consistency constraint
discussed in \cite{Geissbuhler:2013uka}, is important. 

From the point view of a metric algebroid, the gauge consistency constraint can be discussed after
solving the pre-Bianchi identity, i.e., we have to represent the fluxes in terms of the fundamental fields
such as generalized dilaton and generalized vielbein. This opens up a number of possibilities, as we will indicate below, however, the detailed study of them is beyond the scope of this paper. Therefore, in this section we show how 
our formulation produces Bianchi identities and consistency constraint corresponding to the standard DFT case.

\subsection{Closure on $E$}

For the closure of the generalized Lie derivative in the present formulation,
we have to require $\GL(a,b,c)$ in (\ref{WeakmasterDGO1}) to vanish.
Note that vanishing of (\ref{WeakmasterDGO2}) on the Clifford bundle follows.
Thus, we require the following closure condition which corresponds to the weak master equation in the supermanifold approach
\be
\{\{\{\DGO^2,a\},b\},c\}=0~.\label{pClosureconditionofDGOeq}
\ee
Explicit evaluation of $\DGO^2$ yields
\bea
4\DGO^2&=&
\left(\gamma^A\partial_A-{1\over12}F_{ABC}\gamma^{ABC}- \half F_A\gamma^A \right)^2
\cr&=&\partial_A\partial^A
-{1\over24}F_{ABC}F^{ABC}-\half (\rho(E^A)F_A)-F^A\rho(E_A)+{1\over4}F_AF^A
\cr&&
-\left(
{1\over4}\rho(E_A)F_{A'B'C'}\eta^{AA'}+\half\GL'(E_{B'},E_{C'})^A\partial_A+{1\over4} (\rho(E_{[B'})F_{C']}
-F^AF_{AB'C'})\right)\gamma^{B'C'}
\cr&&
-{1\over12}\left(\rho(E_B)F_{CB'C'}
-
{3\over4}F_{ABC}F_{A'B'C'}\eta^{AA'}\right)\gamma^{BCB'C'}.
\eea
We get the explicit form of closure constraint as follows
\bea
0&=&\{\{\{\DGO^2,a\},b\},c\}
\cr&=&-\frac{1}{6}a^Ab^Bc^C\left(\rho(E_{[A})F_{BCD]}-\frac{3}{4}F_{E[AB}F^E{}_{CD]}\right)\gamma^D
\cr&&+\frac{1}{2}\eta_{A[B}\phi'_{CD]}{}^E(\rho(E_E)a^A)b^Bc^C\gamma^D
\cr&&-\frac{1}{2}\eta_{B[C}\phi'_{D]A}{}^Ea^A(\rho(E_E)b^B)c^C\gamma^D
\cr&&+\frac{1}{2}\eta_{CD}\phi'_{AB}{}^Ea^Ab^B(\rho(E_E)c^C)\gamma^D
\cr&&-\left((\rho(E_C)a^{[B|})(\rho(E^C)b^{|D]})c_B-(\rho(E_C)a^{B})(\rho(E^C)c^{D})b_B\right)\gamma_D.
\eea
This condition is understood as a constraint for the structure function $F_{ABC}$
and section $\Gamma(E)$. In principle, we can seek for the solution where $F_{ABC}$ and $\phi'_{ABC}$ satisfy a relation with the coefficients $a^A,b^A,c^A$ and their derivatives. However, for application to DFT in mind, we are interested
in the case where the basis of $\Gamma(E)$ satisfies
the condition
\be
\{\{\{\DGO^2,\gamma_A\},\gamma_B\},\gamma_C\}=0~.\label{pStrongmasterequation}
\ee
Then, we get the Bianchi identity for $F_{ABC}$, i.e., 
\be
\phi_{ABCD}=0~.\label{pphiBianchiidentityforF}
\ee 
where $\phi_{ABCD}$ is given in (\ref{Leibnizidentitytensor}).
For the other terms to vanish we require
\be
\{\{\{\DGO^2,\gamma_A\},\gamma_B\},c\}=0~,\label{pLprimeinDGO}
\ee 
for a set of $c\in\Gamma(E)$ satisfying the following constraint on the coefficients
\be
(\rho(E_C)a^{[B})(\rho(E^C)b^{D]})c_B-(\rho(E_C)a^{B})(\rho(E^C)c^{D})b_B=0~.\label{pClosureconstraint}
\ee
From (\ref{pLprimeinDGO}) it follows $\CL'(E_A,E_B)(c^C)=0$, i.e., by (\ref{pLprimestructurefunction})
\be
\phi'_{AB}{}^E\rho(E_E)(c^C)=0~.\label{pLprimecondition1}
\ee
From (\ref{pClosureconstraint}), which we call the closure constraint, we obtain a restriction on the space of sections $\Gamma(E)$ and we denote this subset as $\Gamma(E)_{ccE}$.

To summarize, the closure of the generalized Lie derivative requires the vanishing of the condition 
(\ref{pClosureconditionofDGOeq}) which we call closure constraint. If we require that the square of the DGO is a function, of course, this condition is satisfied. 
However, here we consider that this condition restricts the structure functions and the space of 
sections in $\Gamma(E)$, like the weak master equation in the supermanifold approach.  

To apply the above formalism to DFT, we require that the basis of $\Gamma(E)$ satisfies the closure condition (\ref{pClosureconditionofDGOeq}),
then this condition implies the Bianchi identity for $F_{ABC}$ and defines $\Gamma(E)_{ccE}$ via the closure constraint.
In standard DFT, the generalized anchor is applied on (\ref{pStrongmasterequation}), then we obtain the constraint
(\ref{pClosureconstraint}) where the coefficients of $a,b,c$ are replaced by the components of the vielbein, which is equivalent to the constraint given in \cite{Geissbuhler:2013uka}\cite{Aldazabal_2013}.

Note that, as we discussed, the Bianchi identity $\phi_{ABCD}=0$ depends on the choice of the basis. However, together with the
conditions (\ref{pClosureconstraint}) and (\ref{pLprimecondition1}), the covariance w.r.t. the rotation of the local frame is recovered.
Note also that we do not get the Bianchi identity for $F_A$ from closure condition (\ref{pClosureconditionofDGOeq}),
which we postpone to the next section.

\subsection{Closure on $\mathbb{S}$}
We have defined the generalized Lie derivative on $\mathbb{S}$ (\ref{GLonS}) with which we can require the closure condition of the generalized Lie derivative on $\mathbb{S}$ as
\be
\{\{\DGO^2,a\},b\}\chi=0~,\label{pClosureconditiononS}
\ee
where $a,b\in\Gamma(E),\chi\in\Gamma(\mathbb{S})$.
Similar to the closure condition on $\Gamma(E)$, eq. (\ref{pClosureconditiononS}) is too strong on arbitrary elements $\chi$.
Therefore, we interpret it as a restriction on $\Gamma({\mathbb{S}})$.
We define a subspace $\Gamma(\mathbb{S})_{cc}\subset\Gamma(\mathbb{S})$ and
$\Gamma(E)_{cc}\subset\Gamma(E)$ whose elements
satisfy the above closure condition. Note that by $\Gamma(E)_{cc}$, we can also define $\Gamma(Cl(E))_{cc}$.
For consistency, $\Gamma(\mathbb{S})_{cc}$ must be a representation of $\Gamma(Cl(E))_{cc}$,
i.e.,
\be
^\forall a\in\Gamma(Cl(E))_{cc},\ ^\forall\chi\in\Gamma(\mathbb{S})_{cc}\Longrightarrow a\chi\in\Gamma(\mathbb{S})_{cc}~.
\ee
This requires
\be
\{\{\DGO^2,a\},b\}\chi=0\Longrightarrow\{\{\DGO^2,a\},b\}c\chi=0~.
\ee
where $a,b,c\in\Gamma(E)_{cc}$ and $\chi\in\Gamma(\mathbb{S})_{cc}$. 
Therefore, $\{\{\{\DGO^2,a\},b\},c\}\chi=0$ and
since there are no derivatives acting on $\chi$ 
we get  the closure condition on $\Gamma(E)$ as
\be
\{\{\{\DGO^2,a\},b\},c\}=0~.
\ee
This condition is equivalent to the one for $\Gamma(E)_{ccE}$ defined in the previous section.

The explicit form of the closure condition for $\chi$ is
\bea
0&=&\{\{4\DGO^2,a\},b\}\chi
\cr&=&\Bigl(\frac{1}{4!}(4\rho(E_{[A})F_{BCD]}-3F_{E[AB}F^E{}_{CD]})a^Ab^B\gamma^{CD}
\cr&&+2\phi'_{AB}{}^Ea^Ab^B\partial_E
\cr&&+2(\rho(E_C)F^C{}_{AB}+\rho(E_{[B})F_{C]}-F^CF_{CAB})a^Ab^B
\cr&&-\frac{1}{2}\eta_{A[B}\phi'_{CD]}{}^E(\rho(E_E)a^A)b^B\gamma^{CD}
\cr&&+\eta_{BD}\phi'_{AC}{}^Ea^A(\rho(E_E)b^B)\gamma^{CD}
\cr&&+2\eta_{AC}\eta_{BD}(\rho(E^E)a^A)(\rho(E_E)b^B)\gamma^{CD}
\cr&&+2\eta_{AB}\left((\rho(E^C)-F^C)(\rho(E_C)a^A)\right)b^B
\cr&&+2\eta_{AB}(\rho(E^E)a^A)(\rho(E_E)b^B)
\cr&&+\phi'_{AB}{}^Ea^A(\rho(E_E)b^B)
\cr&&+4\eta_{AB}(\rho(E^E)a^A)b^B\partial_E\Bigr)\chi~.
\eea
Here, we do not solve this condition in full generality. Instead, we give one example which connects to the standard DFT.
Assume that $\gamma_A$ is a solution of the closure condition as in the discussion 
on the closure on $\Gamma(E)$, i.e., $\gamma_A=E)_A$.
In this way we can get the Bianchi identity for $F_A$ as follows.
The same discussion as in the previous section applies which gives us the Bianchi identity for $F_{ABC}$
and the closure condition on $\Gamma(E)$, i.e., 
(\ref{pphiBianchiidentityforF}), (\ref{pClosureconstraint}) and (\ref{pLprimecondition1}).
Using these equations, the closure constraint on $\Gamma(\mathbb{S})$ reads
\bea
0&=&\{\{4\DGO^2,a\},b\}\chi_0
\cr&=&\Bigl(2\phi'_{AB}{}^Ea^Ab^B\partial_E
\cr&&+2(\rho(E_C)F^C{}_{AB}+\rho(E_{[B})F_{C]}-F^CF_{CAB})a^Ab^B
\cr&&+2\eta_{AC}\eta_{BD}(\rho(E^E)a^A)(\rho(E_E)b^B)\gamma^{CD}
\cr&&+2\eta_{AB}\left((\rho(E^C)-F^C)(\rho(E_C)a^A)\right)b^B
\cr&&+2\eta_{AB}(\rho(E^E)a^A)(\rho(E_E)b^B)
\cr&&+4\eta_{AB}(\rho(E^E)a^A)b^B\partial_E\Bigr)\chi_0~.\label{closureonSgeneral}
\eea
This condition is understood as a constraint on the structure function $F_{A}$
and the section $\Gamma(\mathbb{S})$.

To obtain the Bianchi identity, we 
consider the special case where the base satisfies the closure constraint on $\Gamma(\mathbb{S})$.
Taking $a=\gamma_A,b=\gamma_B\in\Gamma(E)_{cc}$,
we obtain
\bea
0&=&\{\{4\DGO^2,\gamma_A\},\gamma_B\}\chi
\cr&=&2\left(\phi'_{AB}{}^E\partial_E+(\rho(E_C)F^C{}_{AB})+(\rho(E_{[B})F_{C]})-F^CF_{CAB}\right)\chi~.
\label{Bianchi_FA_partial}
\eea
Then we require the closure also for general elements $a,b$. First, taking $b=\gamma_B$ in (\ref{closureonSgeneral}), we get the following relation
\be
\Bigl(2\eta_{AB}\left((\rho(E^C)-F^C)\rho(E_C)a^A\right)+4\eta_{AB}\left(\rho(E^E)a^A\right)\partial_E\Bigr)\chi=0~.
\ee
Using this equation we finally obtain for general elements $a,b$
\be
\Bigr(2\eta_{AC}\eta_{BD}(\rho(E^E)a^A)(\rho(E_E)b^B)\gamma^{CD}+2\eta_{AB}(\rho(E^E)a^A)(\rho(E_E)b^B)\Bigr)\chi=0~.
\ee
Thus, the restriction of the vector bundle
$\Gamma(E)_{ccE}\subset\Gamma(E)$ is not enough to satisfy the closure condition on $\Gamma(\mathbb{S})$,
i.e., $\Gamma(E)_{ccE}\subset\Gamma(E)_{cc}$. In standard DFT, this condition is satisfied by the strong
constraint $(\rho(E_C)a^A)(\rho(E^C)b^B)=0$.

Furthermore, assuming that $\Ket{0}\in\Gamma(\mathbb{S})_{cc}$, we get the Bianchi identity for $F_A$
by the equation (\ref{Bianchi_FA_partial}),
\be
(\rho(E_C)F^C{}_{AB})+(\rho(E_{[B})F_{C]})-F^CF_{CAB}=0~,
\ee
and
\be
\phi'_{AB}{}^E\rho(E_E)f=0~,
\ee
where $f$ is a function in $\Gamma(E)_{cc}$.
In this case, $\Gamma(\mathbb{S})_{cc}$ becomes
\be
\Gamma(\mathbb{S})_{cc}=\{O\Ket{0}|O\in\Gamma(Cl(E))_{cc}\}~.
\ee
To summarize, in the case where $\gamma_A\in\Gamma(E)_{cc}$ and $\Ket{0}\in\Gamma(\mathbb{S})_{cc}$,
 the closure on $\Gamma(\mathbb{S})$ requires
\bea
\rho(E_{[A})F_{BCD]}-\frac{3}{4}F_{E[AB}F^E{}_{CD]}&=&0~,
\cr\phi'_{AB}{}^E\rho(E_E)(c^C)&=&0~,
\cr\rho(E_C)F^C{}_{AB}+\rho(E_{[B})F_{C]}-F^CF_{CAB}&=&0~,
\cr\phi'_{AB}{}^E\rho (E_E)f&=&0~,
\cr\eta_{AB}(\rho(E^C)-F^C)\rho(E_C)a^A&=&0~,
\cr(\rho(E^E)a_{[A})(\rho(E_E)b_{B]})&=&0~,
\cr\eta_{AB}(\rho(E^E)a^A)(\rho(E_E)b^B)&=&0~.
\eea
In this way, we obtain the Bianchi identities for $F_A$ and $F_{ABC}$.
In standard DFT, the Bianchi identity is solved by imposing the strong constraint.
In this case $\eta_{MN}$ is constant and a solution of the Bianchi identity is given by 
$F_{ABC}=\CF_{ABC}$ in (\ref{pStandardDFTflux}) and $F_A=\CF_A$ in (\ref{pStandardDFTfluxwithsingleflux}).

\section{Conclusion and discussion}

In this paper, after giving a brief survey on the algebraic structure of a metric algebroid, we analyzed the properties of the structure functions relating to DFT. By requiring independence of the choice of the local bases, we found that a pre-Bianchi identity can be obtained as a completion of the  Bianchi identities.
As a result we obtain the map $\tilde\phi$ the vanishing of which is a pre-Bianchi identity. 

The structure of metric algebroid is fixed by structure functions defining backet and anchor on the vector bundle $E$.
On the other hand, to formulate the geometrical objects on the vector bundle such as torsion and curvature 
we have to introduce a connection. The metric algebroid does not fix all components of these objects. However, we can obtain the pre-Bianchi identity in terms of generalized torsion and a covariant generalized curvature, eq. (\ref{pCovariantBianchiinpreCA}), where this curvature is also a base independent completion of the generalized curvature of standard DFT w.r.t. the metric algebroid. 

Another aim of this paper was to find the origin of the Bianchi identity including the dilaton in DFT, which we could not achieve in the QP manifold approach. It turns out that rewriting the Jacobi identities on $T\mathbb{M}$ in terms of the structure functions of the metric algebroid, there is a freedom which allows us to introduce the dilaton. 
From this point of view, the flux is an ambiguity which is not fixed by the structure function of the metric algebroid.
The flux including the dilaton became clear when we considered the Dirac generating operator approach. 
In summary we can say that from the metric algebroid point of view these pre-Bianchi identities select a sub-class to which DFT belongs.

In the second part of this paper, we gave a formulation of DFT using the Dirac generating operator (DGO). Unlike in generalized geometry, we did not require the square of the DGO to be a function. This relaxation of the condition on the DGO lead us to the structure of a metric algebroid.
In this setting the DGO is the fundamental object and its square 
contains differential operators in general. This investigation gave several new insights.

From the square of this DGO with derivative terms covariantly subtracted we derived the pre-Bianchi identities with which we have characterized the metric algebroid underlying DFT.
After the subtraction, the square of the DGO contains three contributions, and requiring the result to be a 
scalar function we obtain both pre-Bianchi identities, i.e. (\ref{explicitformofpreBianchi}) and (\ref{pBianchiofOneIndex1}), and the scalar part becomes the scalar of the covariant generalized curvature (\ref{pCovariantcurvaturescalar1}).   
This procedure results in a generalized Lichnerowicz formula.
Thus, the condition for the pre-Bianchi identities to hold is equivalent to the condition that the generalized Lichnerowicz formula is satisfied. 
Given a metric algebroid, there is an ambiguity in the DGO, and this freedom allows to introduce the dilaton into the 
 structure function $F_A$.

To obtain the action, we introduce a Riemann structure by a splitting of the vector bundle into positive and negative subbundle. By using the corresponding projection we obtain the projected generalized Lichnerowicz formula, which is proportional to the projected generalized scalar curvature under the 
pre-Bianchi identity.
Then, we propose an action for DFT in terms of the projected Lichnerowicz formula.
To formulate the measure of the action,
we introduced the inner product of the pure spinor $\mu=\braket{0|0}$ which is $O(D,D)$ invariant. We could interpret $\mu$ as the measure $dX^1\wedge\cdots\wedge dX^{2D}e^{-2d}$
 in the standard DFT choosing the representation. However, $\mu$ may not be in $\wedge T\mathbb{M}$ in general.

{ \bf Remarks on generalized supergravity equations (GSE)}

Recently, 
a generalization of supergravity, originally proposed in \cite{ARUTYUNOV2016262,Tseytlin_2016},  is discussed by several authors as a possibility to modify the supergravity equations to a more general set of field 
equations in the context of integrable deformations, keeping consistency with superstring \cite{Delduc_2014,Kawaguchi_2014}.
These integrable deformations are considered to be closely related to non-Abelian T-duality transformations \cite{DELAOSSA1993377,GIVEON1994173,ALVAREZ199471} and also to Poisson-Lie T-duality \cite{KLIMCIK1995455,KLIMCIK1996116}. 

One way to obtain the generalized supergravity equations (GSE) in DFT which fits to the approach given here  
 is to consider a modification of the field representation of the structure function $F_A$. 
 This modification is possible due to the ambiguity $X$ in the divergence compatible with the splitting $V^{\pm}$, which is discussed in \cite{Severa:2018aa} in the context 
 of generalized geometry.

While in standard DFT the structure function is represented by $F_A=\CF_A$, by the ambiguity in the divergence 
the structure function $F_A$ can include a generalized Killing vector $X$ as follows:
\be
F_A=\CF_A+X_A
\ee
where $X_A$ satisfies
\be
\CL_X\CH=0~.
\ee
where $\CH$ is a generalized metric (\ref{pGeneralizedmetrictensor}) in the appendix.
Since the combination of the pre-Bianchi identity (\ref{explicitformofpreBianchi}) is covariant but not necessarily zero,
we may extend it by a covariant term as
\be
{{\rho(E_{C})F_{AB}{}^C}}
-{{F_{AB}{}^{C'}F_{C'}}}
+{{\rho(E_{[A})(F_{B]}})}
-\rho(E_{C})\phi'_{AB}{}^C+\phi'_{AB}{}^{C'}F_{C'}=E_A^ME_B^N\partial_{[M}X_{N]}~,
\ee
where the r.h.s. is an additional term corresponding to a derivative: $\Gamma(T\mathbb{M})\rightarrow\Gamma(T\mathbb{M})\wedge\Gamma(T\mathbb{M})$.
The structure function $F_A=\CF_A+X_A$ is a solution of this covariant equation.
This means that $X$ can be interpreted as a freedom in the pre-Bianchi identity.

Furthermore, if $X$ satisfies $\partial_{[M}X_{N]}=0$, then the pre-Bianchi identity for $F_A$ becomes zero, which we required in this paper to characterize the metric algebroid for DFT.
The simplest solution of this condition is
 $X_M=$ constant. In this case, the ambiguity $X$ of the structure function $F_A$
 becomes a constant Killing vector, which is used as an ansatz for the dilaton to obtain the GSE from DFT with non-standard section \cite{10.1093/ptep/ptx067}.

\section*{Acknowlegments}
The authors would like to thank G. Aldazabal, P. \v{S}evera and K. Yoshida
for stimulating discussions and lectures.
We also would like to thank T. Kaneko, S. Sekiya, S. Takezawa and in particular N. Ikeda for valuable discussions.  
S.W. is supported by the JSPS Grant-in-Aid for Scientific Research (B) No.18H01214.

\appendix
\renewcommand{\theequation}{\thesection.\arabic{equation}}

\section{Spin bundle}

\subsection{Spin bundle of $E$}
We define the spin bundle $\mathbb{S}$ as a module of $\Gamma(Cl(E))$.
We construct a explicit form of $\mathbb{S}$ by a pure spinor $\Ket{0}$.
The definition of a pure spinor $\Ket{0}$ is a element on $\mathbb{S}$ which vanishes
when operators on a maximal isotropic subspace of $E\subset\mathbb{S}$ act on $\Ket{0}$.
In this paper, $E$ is split into positive and negative definite D-dimensional subspace $V^+,V^-$,
i.e., $E=V^+\oplus V^-$. Then, $E$ can be written by D-dimensional isotropic subspaces $L_1,L_2$,
i.e., $E=L_1\oplus L_2$. We define a basis $l^a,l_a$ of $L_1,L_2$, respectively,
\be
l^a\in L_1~,~~l_a\in L_2~,
\ee
where
\be
\braket{l^a,l^b}=\braket{l_a,l_b}=0~,~~\braket{l_a,l^b}=\delta_a^b~.
\label{def_basisl}
\ee
The Clifford action satisfies $\{a,b\}=2\braket{a,b}$, see (\ref{pCliffordbracketonframe}),
and the pure spinor is defined by
\be
l_a\Ket{0}=0 ~~. \label{def_purespinor}
\ee
Using these operators and a pure spinor, we can define the space of sections of $\mathbb{S}$ in this representation
 as follows
\be
\Gamma(\mathbb{S})=\Bigl\{C\Ket{0}+\sum_{n=1}^{D}C_{a_1\cdots a_n}l^{a_1}\cdots l^{a_n}\Ket{0}|C,C_{a_1\cdots a_n}\in C^\infty(\mathbb{M}) \Bigr\}~.
\ee

\subsection{Spin bundle of $V^+$}
As in the previous section, we consider a spin bundle $\Gamma(\mathbb{S}^+)$
 as a module of $\Gamma(Cl(V^+))$. The vector bundle $E$ can be written by the positive and 
 negative subbundles $E=V^+\otimes V^-$. Then, the Clifford bundles $Cl(V^\pm)$ can be defined by a bracket 
 $\{-,-\}$ for $a,b\in V^\pm\subset Cl(V^\pm)$,
\be
\{a,b\}=2\braket{a,b}.
\ee
Now, assume that $V^+$ admits a Clifford module $\mathbb{S}^+$, i.e.,
 $V^+$ can be split into $D/2$-dimensional isotropic subbundles $L_1$ and $L_2$ where $V^+=L_1\oplus L_2$. We define a basis $l^a,l_a$ of $L_1,L_2$, respectively.
A pure spinor $\Ket{0}$ is defined by
\be
l_a\Ket{0}=0~.
\ee
Using this pure spinor, we can construct the spin bundle $\Gamma(\mathbb{S}^+)$ as
\be
\Gamma(\mathbb{S}^+)=\Bigl\{C\Ket{0}+\sum_{n=1}^{D/2}C_{a_1\cdots a_n}l^{a_1}\cdots l^{a_n}\Ket{0}|C,C_{a_1\cdots a_n}\in C^\infty(\mathbb{M}) \Bigr\}~.
\ee

\section{Divergence}

\subsection{Divergence in DFT}

Given a Courant algebroid $E$, the divergence 
is defined \cite{AlekseevDGO, Garcia-Fernandez:2016aa, Severa:2018aa} 
as a map  $div:\Gamma(E)\rightarrow C^\infty(\mathbb{M})$, for $a\in\Gamma(E)$ and $f\in C^\infty(\mathbb{M})$ s.t. 
\be
div(fa)=\rho(a)f + f div(a)~.\label{pdivergencerelation}
\ee
We apply the same definition to the metric algebroid. 

For a given $E$-connection, the corresponding divergence $div_\nabla$ is given by
\be
div_\nabla a=\braket{\nabla^E_{A}a , E^A}~,
\label{pDef:divnabla}
\ee
where $div_\nabla$ satisfies the relation (\ref{pdivergencerelation}), i.e.,
\bea
div_\nabla f a= \braket{\nabla^E_{A}f a , E^A}
=\braket{(\rho(E_A)f)a, E^A}+\braket{f\nabla^E_{A}a , E^A}
=\rho(a)f+f div_\nabla a~.
\eea
Note that due to the property of the covariant derivative, the summation over the basis does not depend on 
the choice of the basis.
Using the local basis, we obtain
\bea
div_\nabla a
&=& \rho(E_A)a^A-W_{B}{}^{BA}a_A~.\label{pLocalformofdivnabla}
\eea

Considering an arbitrary divergence $div$ which satisfies the relation (\ref{pdivergencerelation}), 
the difference between $div_\nabla$ and $div$ is a $C^\infty(\mathbb{M})$-linear function 
\be
div(fa)-div_\nabla(fa)=f(div(a)-div_\nabla(a))~.
\ee
Thus, the difference is characterized by $U\in\Gamma(E)$ as $div-div_\nabla=-\braket{U,-}$, i.e., 
 the general form of the divergence has an ambiguity $U\in\Gamma(E)$ which can be written as
\be
div(a)=div_\nabla(a)-\braket{U,a}~.
\label{general_div_onE}
\ee

\subsection{Laplace operator for general representation}

We define the divergence $div$ on an arbitrary vector bundle $L$ in the same way as (\ref{pdivergencerelation}).
A divergence on $\Gamma(L)$ is a map $div:\Gamma(E)\otimes\Gamma(L)\rightarrow\Gamma(L)$ s.t.
\be
div(fa\otimes\sigma)=(\rho(a)f)\sigma+fdiv(a\otimes\sigma)~,\label{pdivergencerelationonL}
\ee
where $f\in C^\infty(M),a\in\Gamma(E),\sigma\in\Gamma(L)$.

Assume that a connection $\nabla^L$ acting on $L$ exists, i.e.,
$\nabla^L:\Gamma(E)\times\Gamma(L)\rightarrow\Gamma(L)$, s.t.
\bea
\nabla^L_{fa}(\sigma)&=&f\nabla^L_a\sigma\ ,
\cr\nabla^L_a(f\sigma)&=&(\rho(a)f)\sigma+f\nabla^L_a\sigma~,
\eea
where $f\in C^\infty(\mathbb{M}),a\in\Gamma(E),\sigma\in\Gamma(L)$.
Then, we can define a connection $\nabla^{E\otimes L}:\Gamma(E)\times \Gamma(E)\otimes\Gamma(L)\rightarrow\Gamma(E)\otimes\Gamma(L)$ by Leibniz rule,
\be
\nabla_{a}^{E\otimes L}{b}\otimes \sigma=(\nabla^E_ab)\otimes \sigma+b\otimes \nabla^{L}_{a}\sigma~,
\ee
where $a,b\in \Gamma(E),\ \sigma\in\Gamma(L)$. Using this connection, we can define the divergence $div_\nabla:\Gamma(E)\otimes\Gamma(L)\rightarrow\Gamma(L)$ as
\be
div_\nabla=\iota_{E^A}\nabla_{E_A}^{E\otimes L}~,
\ee
where the contraction $\iota_a:\Gamma(E)\otimes \Gamma(L)\rightarrow \Gamma(L)$ is defined by
\be
\iota_a(b\otimes\sigma)=\braket{a,b}\sigma~.
\ee
This divergence satisfies (\ref{pdivergencerelationonL}) as follows
\be
div_\nabla(fa\otimes\sigma)=(\rho(a)f)\sigma+fdiv_\nabla(a\otimes\sigma)~.
\ee
Given an arbitrary divergence $div$, the difference $div-div_\nabla$ satisfies
\be
(div-div_\nabla)(fa\otimes\sigma)=f(div-div_\nabla)(a\otimes\sigma)~.
\ee
This means that $div-div_\nabla$ is a linear map $\Gamma(E)\otimes\Gamma(L)\rightarrow\Gamma(L)$, i.e.,
$^\exists U\in\Gamma( E), \ div-div_\nabla=-\iota_U$.
Thus, an arbitrary divergence can be written by
\be
div=div_\nabla^U:=div_\nabla-\iota_U~.
\ee
To construct a Laplace operator on $\Gamma(L)$ we define a connection $\nabla^{L}:\Gamma(L)\rightarrow \Gamma(E)\otimes\Gamma(L)$ such that
\be
\iota_a\nabla^{L}\sigma=\nabla^{L}_a\sigma~.
\ee
An explicit form of this connection is
\be
\nabla^L\sigma=E^A\otimes\nabla^{L}_{E_A}\sigma~.
\ee
Using these operators a Laplace operator $\Delta:\Gamma(L)\rightarrow\Gamma(L)$ can be defined as
\bea
\Delta^{L}&=&div_\nabla^U\nabla^{L}~.
\eea
These definitions coincide with the divergence (\ref{general_div_onE}) on $E$. We will see that 
the standard Laplacian on functions is also included.  
In the following we show concrete forms of the divergence and the Laplace operator 
for $\Gamma(L)=C^\infty(\mathbb{M}), 
\Gamma(\mathbb{\mathbb{S}})$ and $\Gamma(\mathbb{S}^+)$, respectively.

\subsubsection{Laplace operator on $C^\infty(\mathbb{M})$}
In the case where $\Gamma(L)=C^\infty(\mathbb{M})$, we obtain
\bea
\nabla^{L}_af&=&\rho(a)f\ ,
\\\nabla_a^{E\otimes L}&=&\nabla_a^E\ ,
\\ div_\nabla^Ua&=&(\iota_{E^A}\nabla^E_{E_A}-\iota_U)a
\cr&=&\braket{E^A,\nabla_{E_A}a}-\braket{U,a} \ ,
\\\Delta^Lf&=&div_\nabla^U\nabla^Lf
\cr&=&div_\nabla^U(E^A\rho(E_A)f)
\cr&=&(\iota_{E^B}\nabla^E_{E_B}-\iota_U)(E^A\rho(E_A)f)
\cr&=&\rho(E^A)(\rho(E_A)f)+W_B{}^{AB}\rho(E_A)f-\rho(U)f
\cr&=&(\rho(E^A)-W_B{}^{BA}-U^A)(\rho(E_A)f)\ ,
\eea
where $\nabla^E_{E_A}E_B=W_{AB}{}^CE_C, a\in\Gamma(E), f\in C^\infty(\mathbb{M})$.
This divergence $div_\nabla^U$ is equal to the divergence (\ref{general_div_onE}). For $U=0$ 
the Laplace operator is the standard Laplacian on $C^\infty(\mathbb{M})$.

\subsubsection{Laplace operator on $\mathbb{S}$}
In the case where $L=\mathbb{S}$, we obtain
\bea
\nabla^{{L}}_a&=&\nabla^\mathbb{S}_a\ ,
\\\nabla_a^{E\otimes L}&=&\nabla_a^{E\otimes \mathbb{S}}\ ,
\\ div_\nabla^U&=&\iota_{E^A}\nabla^{E\otimes \mathbb{S}}_{E_A}-\iota_U\ ,
\label{pAxiomDivergenceonSpinRepApp}
\\\Delta^L\chi&=&div_\nabla^U\nabla^\mathbb{S}\chi
\cr&=&(\nabla^\mathbb{S}_{E^A}-W_B{}^{BA}-U^A)\nabla^\mathbb{S}_{E_A}\chi\ ,
\eea
where $a\in\Gamma(E)$ and $\chi\in\Gamma(\mathbb{S})$.

\subsubsection{Laplace operator on $\mathbb{S}^+$}
Considering the case where $L=\mathbb{S}^+$,
we can define a Laplace operator on $\mathbb{S}^+$.
Since $\mathbb{S}^+$ has only a $Cl(V^+)$ action, there is no $O(D,D)$ transformation but only $O(D)$.
Thus, we can construct a connection on $\mathbb{S}^+$ which has only $O(D)$ covariance
using the E-connection $\nabla^E_{E_A}E_B=W_{AB}{}^CE_C$.
An explicit form of this connection $\nabla^{\mathbb{S}^+}:\Gamma(E)\times\Gamma(\mathbb{S}^+)\rightarrow\Gamma(\mathbb{S}^+)$ in the basis $E_A$ is
\be
\nabla^L_{E_A}=\nabla_{E_A}^{\mathbb{S}^+}=\partial_A-\frac{1}{4}W_{Aab}\gamma^{ab}~,
\ee
where $\gamma^a$ is the basis of $Cl(V^+)$ and $\partial_A$ is a zero connection on $\mathbb{S}^+$,
i.e., for $f\in C^\infty(\mathbb{M})$ and the pure spinor $\Ket{0}$ on $\mathbb{S}^+$,
\bea
\{\partial_A,f\}&=&\rho(E_A)f~,
\cr\{\partial_A,\gamma^a\}&=&0~,
\cr\partial_A\Ket{0}&=&0~.
\eea
Using this connection, we can define the connection $\nabla^{E\otimes\mathbb{S}^+}$, the divergence and the Laplace operator as
\bea
\nabla_a^{E\otimes L}&=&\nabla_a^{E\otimes \mathbb{S}^+}\ ,
\\ div_\nabla^U&=&\iota_{E^A}\nabla^{E\otimes \mathbb{S}^+}_{E_A}-\iota_U\ ,
\\\Delta^L\chi^+&=&div_\nabla^U\nabla^\mathbb{S^+}\chi^+
\cr&=&(\nabla^{\mathbb{S}^+}_{E^A}-W_B{}^{BA}-U^A)\nabla^{\mathbb{S}^+}_{E_A}\chi^+\ ,
\eea
where $a\in\Gamma(E)$ and $\chi^+\in\Gamma(\mathbb{S}^+)$.

\subsection{Compatible divergence}\label{compatibledivergenceappend}
In this paper, a metric algebroid is defined on a vector bundle $\Gamma(E)=\Gamma(V^+)\oplus\Gamma(V^-)$,
where $V^{+},V^-$ are positive and negative definite subbundles, respectively.
In order to discuss a compatible divergence with $V^{\pm}$ as in \cite{Severa:2018aa},
 we prepare some structures as follows.
First, we reformulate the projection $E\rightarrow V^{\pm}$ in (\ref{pDefprojectionpm}),
defining two tensors $\eta$ and $\CH$ on $E\otimes E$,
\bea
\eta&=&\eta_{AB}E^A\otimes E^B~,\\
\CH&=&\CH_{AB}E^A\otimes E^B~, \label{pGeneralizedmetrictensor}
\eea
where $\eta_{AB}$ and $\CH_{AB}$ are defined by the inner product $\braket{-,-}$
and the generalized metric (\ref{pGeneralizedMetricasmap}), respectively.
The projection tensor $\CP^\pm\in E\otimes E$ is
\be
\CP^\pm=\frac{1}{2}(\eta\pm \CH) ~.
\ee
Using the above tensor, we can write the projection $E\rightarrow V^{\pm}$ as
\be
\CP^\pm(a)=\iota_a\CP^\pm~,
\ee
where $a\in E$.
The generalized Lie derivative on $\Gamma(E)\otimes \Gamma(E)$ is defined by the Leibniz rule:
\be
\CL_a(b\otimes c)=[a,b]\otimes c+b\otimes[a,c]~.
\ee
Then, the generalized Lie derivative satisfies the
Leibniz rule also for the contraction $\iota_a$
\bea
\CL_d\iota_a(b\otimes c)&=&\CL_d(\braket{a,b}c)
\cr&=&\rho(d)(\braket{a,b})c+\braket{a,b}[d,c]
\cr&=&\braket{[d,a],b}c+\braket{a,[d,b]}c+\braket{a,b}[d,c]
\cr&=&\iota_{[d,a]}b\otimes c+\iota_a[d,b]\otimes c+\iota_ab\otimes [d,c]
\cr&=&\iota_{[d,a]}b\otimes c+\iota_a\CL_d(b\otimes c)~,
\eea
where $a,b,c,d\in\Gamma(E)$.

From this definition we can show
\be
\CL_a\eta=0~.\label{Liederivativeofmetric}
\ee
To see this we calculate $\CL_ab$ 
\bea
\CL_ab&=&\CL_a\iota_b\eta
\cr&=&\iota_{[a,b]}\eta+\iota_b\CL_a\eta
\cr&=&[a,b]+\iota_b\CL_a\eta~,
\eea
where we used $\iota_a\eta=a$. Since $b\in\Gamma(E)$ is an arbitrary element, this means that (\ref{Liederivativeofmetric})
holds.

We define the divergence $div^{com}$ compatible with $V^{\pm}$  as follows:
\be
div^{com}=div_{\nabla^{\phi'}}^{2\partial d}-\iota_X~,
\ee
where $X\in\Gamma(E)$ satisfies
\be
\braket{[X,\iota_a\CP^\pm],\iota_b\CP^\mp}=0~,
\label{compatible_div_X}
\ee
i.e., $div^{com}=div_{\nabla^{\phi'}}^{U}$ for $U=2\partial d+X$.
This condition means that the map $[X,-]:\Gamma(E)\rightarrow\Gamma(E)$
is closed on $V^{\pm}$, respectively, i.e., $[X,-]:\Gamma(V^{\pm})\rightarrow\Gamma(V^{\pm})$.
Furthermore, under the pre-Bianchi identity, it holds 
\be
div^{com}(E_A)\Big|_{X=0}=-F_A~.\label{pdivandfluxFA}
\ee

The condition for $X\in\Gamma(E)$(\ref{compatible_div_X}) can be written in a simpler form as follows.
\bea
0=\braket{[X,\iota_a\CP^\pm],\iota_b\CP^\mp}&=&\braket{\iota_{[X,a]}\CP^\pm,\iota_b\CP^\mp}+\braket{\iota_a\CL_X\CP^\pm,\iota_b\CP^\mp}
\cr&=&\braket{\pm\frac{1}{2}\iota_a\CL_X\CH,\iota_b\CP^\mp}
\cr&=&\braket{\pm\frac{1}{2}\iota_{(\iota_a\CL_X\CH)}\CP^\mp,\iota_b\CP^\mp}~,
\eea
where we use $\CL_a\eta=0$.
Since $b\in\Gamma(E)$ is an arbitrary element in $\Gamma(E)$, we get
\be
\iota_{(\iota_a\CL_X\CH)}\CP^\mp=0~,
\ee
\be
\iota_a\CL_X\CH=\iota_{(\iota_a\CL_X\CH)}\eta=\iota_{(\iota_a\CL_X\CH)}(\CP^++\CP^-)=0~.
\ee
Thus, $X$ satisfies
\be
\CL_X\CH=0~.
\ee
This condition means that $X$ is a generalized Killing vector.
In the case where we use this compatible divergence for the Laplace operator $\Delta^{\phi'^+}$ in the DFT action,
the structure function $F_A$ becomes
\be
F_A=\phi'_{BA}{}^B+U_A=\phi'_{BA}{}^B+2\rho(E_A)d+X_A~.
\ee

The similar condition has been considered in the generalized geometry in \cite{Severa:2018aa}.
The generalization considered here has an ambiguity. In the generalized geometry,
a special divergence $div_\mu$ is used instead of $div_{\nabla^{\phi'}}^{2\partial d}$.
$div_\mu$ is defined by a $D$-form in $\wedge T^*M$,
\be
div_\mu a=\mu^{-1}\CL_{\rho(a)}\mu~,
\ee
where $\CL$ is the standard Lie derivative.
$div_\mu$ satisfies
\be
div_\mu[a,b]-\rho(a)div_\mu b+\rho(b)div_\mu a=0~,
\ee
where $a,b$ are elements of a Courant algebroid and $[-,-]$ is the bracket of this Courant algebroid.
On the other hand, in DFT, $div_{\nabla^{\phi'}}^{2\partial d}$ does not satisfy such a condition.
However,  $div_{\nabla^{\phi'}}^{2\partial d}$
can be identified with $div_\mu$ in the supergravity frame as follows:
We use the ansatz of the vielbein $E_a=-e_a^mB_{mn}dX^n+e_a^n\partial_n,E^a=e^a_ndX^n$, 
where $e_a^n,B_{mn}$ are identified with the vielbein and the Kalb-Ramond field, respectively, in generalized geometry.
We obtain
\bea
div_{\mu}E_a&=&\partial_ne^n_a-2e_a^n\partial_nd~,
\cr div_{\mu}E^a&=&0~,
\eea
where $\mu=e^{-2d}dX^1\wedge\cdots\wedge dX^D$.
On the other hand, $div_{\nabla^{\phi'}}^{2\partial d}$ becomes
\bea
div_{\nabla^{\phi'}}^{2\partial d}E_a&=&\partial_ne_a^n-2e_a^n\partial_nd~,
\cr div_{\nabla^{\phi'}}^{2\partial d}E^a&=&0~.
\eea
where $E_a=-e_a^mB_{mn}\partial^n+e_a^n\partial_n,E^a=e^a_n\partial^n$ corresponding to $E_A\in TM\oplus T^*M$ and
$\phi'_{AB}{}^C=\Omega^C{}_{AB}$.
Thus, $div_{\nabla^{\phi'}}^{2\partial d}$
can be identified with $div_\mu$ in the supergravity frame in standard DFT.
But the identification of $\mu$ has an ambiguity corresponding to the dilaton shift. In generalized geometry this ambiguity in the choice of $\mu$ is
not important since it changes $X$ by total derivative and thus the condition of the compatible divergence is the same.
On the other hand, in DFT this ambiguity changes the condition of the compatible divergence, since the generalized Killing vector $X$ shifted by the total derivative is not a generalized Killing vector, in general. So, the generalization of the compatible divergence is not unique. Here, this ambiguity is fixed by the condition (\ref{pdivandfluxFA}).

\section{Density of the dilaton}\label{AppendixDilatondensity}

The exponetial of the dilaton $e^{-d}$ is a density with weight ${1\over2}$. 
As we discussed in \S\ref{pGeneralizedLieDerivativesection}, the generalized Lie derivative on $\chi \in \Gamma(\mathbb{S})$ is given by
\be
\CL_a\chi=\{\slashed D,a\}\chi~.
\ee
In this section, we split the spin bundle $\Gamma(\mathbb{S})$ 
into a spin bundle $\Gamma(\mathbb{S}')$ and a line bundle $\Gamma(\Lambda^{\frac{1}{2}})$ which carries the weight $\half$, 
i.e., $\Gamma(\mathbb{S})=\Gamma(\mathbb{S}')\otimes\Gamma(\Lambda^\frac{1}{2})$.
First, using the basis $l^{n}$ defined in (\ref{def_basisl}), we define the spin bundle $\Gamma(\mathbb{S}')$ and the line bundle $\Gamma(\Lambda^\frac{1}{2})$ as follows:
\bea
\Gamma(\mathbb{S})&=&\Gamma(\mathbb{S}')\otimes\Gamma(\Lambda^\frac{1}{2})\ ,
\\\Gamma(\mathbb{S}')&=&\{C\Ket{0'}+\sum_{n=1}^{D}C_{a_1\cdots a_n}l^{a_1}\cdots l^{a_n}\Ket{0'}\ |\ C,C_{a_1\cdots a_n}\in C^\infty(\mathbb{M})\}\ .
\\\Gamma(\Lambda^{\frac{1}{2}})&=&\{f\mu^\frac{1}{2}\ |\ f\in C^\infty(\mathbb{M})\}~,\label{LinebundleLambda}
\eea
where the generalized Lie derivative acts on 
$\Ket{0'}$ and $\mu^\frac{1}{2}$ as
\bea
\Ket{0}&=&\Ket{0'}\otimes\mu^\frac{1}{2},
\\ \CL_a\Ket{0'}&=&\frac{1}{2}\Big(-\half F_{AB}{}^Ca_C+\rho(E_{[A})a_{B]}\Big)\gamma^{AB}\Ket{0'}
\cr&=&\frac{1}{2}\Big(-\frac{1}{2}\phi'_{AB}{}^Ca_C+\partial_{[N}a_{M]}E_A^NE_B^M\Big)\gamma^{AB}\Ket{0'}~,
\\ 
\CL_a\mu^\frac{1}{2}&=&\left(\frac{1}{2}\rho(E_A)a^A-\frac{1}{2}F_Aa^A\right)\mu^\frac{1}{2}
\cr&=&\left(-\rho(a)d+\frac{1}{2}\rho(E_A)a^A-\frac{1}{2}\phi'_{BA}{}^Ba^A\right)\mu^\frac{1}{2}~.
\eea
Note that this definition is consistent with the Leibniz rule, i.e., $\CL_a(\Ket{0'}\otimes\mu^\frac{1}{2})=(\CL_a\Ket{0'})\otimes\mu^\frac{1}{2}+\Ket{0'}\otimes\CL_a\mu^\frac{1}{2}$.

The meaning of this separation of the basis can be understood by considering a inner product on $\Gamma(\mathbb{S}')$ as follows. We assume that the inner product 
$(-,-):\Gamma(\mathbb{S}')\times \Gamma(\mathbb{S}')\rightarrow \Gamma(L')$
exists where $L'$ is a line bundle. 
This inner product satisfies following properties
for $f\in C^\infty(\mathbb{M})$, $\chi_1,\chi_2\in\mathbb{S}'$ and $a\in\Gamma(E)$
\bea
(f\chi_1,\chi_2)&=&(\chi_1,f\chi_2)=f(\chi_1,\chi_2)\ ,
\\(\chi_1,a\chi_2)&=&(a\chi_1,\chi_2)\label{def_dagger}\ ,
\\\delta_a(\chi_1,\chi_2)&=&(\delta_a\chi_1,\chi_2)+(\chi_1,\delta_a\chi_2)\ .
\eea
Note that $a\in\Gamma(E)$ is real section.
In the case where $\chi_1$ and $\chi_2$ are given by $A_1\Ket{0'}$ and $A_2\Ket{0'}$ $(A_1,A_2\in\Gamma(Cl(E)))$, respectively, and denoting $\Bra{0'}A_1^\dagger :=(A_1\Ket{0'},-)$,
the inner product is written as
\be
\Bra{0'}A_1^\dagger A_2\Ket{0'}:=(A_1\Ket{0'},A_2\Ket{0'})\ ,
\ee
where $\dagger$ is defined by $(A^\dagger\chi_1,\chi_2)=(\chi_1,A\chi_2)$ using (\ref{def_dagger}).

Here, we consider the $O(D,D)$ rotation on $\chi\in\Gamma(\mathbb{S}')$ defined by
\be
\delta_{\Lambda}\chi=-\frac{1}{4}\Lambda_{AB}\gamma^{AB}\chi~,
\ee
where $\Lambda_{AB}$ is an anti-symmetric matrix. In particular, we assume that $\gamma_A$ satisfies 
$\gamma_A^\dagger=\gamma_A$.
From the definition of the inner product, we can see that it is 
invariant under the $O(D,D)$ rotation, i.e.,
\be
\delta_\Lambda(\chi_1,\chi_2):=(\delta_\Lambda\chi_1,\chi_2)+(\chi_1,\delta_\Lambda\chi_2)=0\ .
\ee
Since the generalized Lie derivative on $\Ket{0'}$ is a rotation of $\Gamma(\mathbb{S}')$, 
the inner product of $\Ket{0'}$ is invariant, i.e.,
\bea
\CL_a\braket{0'|0'}&=&\left(\frac{1}{2}\Big(-\frac{1}{2}\phi'_{AB}{}^Ca_C+\partial_{[N}a_{M]}E_A^NE_B^M\Big)\gamma^{AB}\Ket{0'},\Ket{0'}\right)
\cr&&+\left(\Ket{0'},\frac{1}{2}\Big(-\frac{1}{2}\phi'_{AB}{}^Ca_C+\partial_{[N}a_{M]}E_A^NE_B^M\Big)\gamma^{AB}\Ket{0'}\right)
\cr&=&0~.
\eea
We can identify $\braket{0'|0'}$ with a constant scalar, since they have the same transformation property. 
Similarly, we can also identify $\braket{0'|\gamma_{A_1}\cdots\gamma_{A_n}|0'}$ with a constant
scalar, since the generalized Lie derivative on $\gamma_{A_1}\cdots\gamma_{A_n}\Ket{0'}$ is given by
$\frac{1}{2}\Big(-\frac{1}{2}\phi'_{AB}{}^Ca_C+\partial_{[N}a_{M]}E_A^NE_B^M\Big)\gamma^{AB}\gamma_{A_1}\cdots\gamma_{A_n}\Ket{0'}$.

Furthermore, we can show that the line bundle $\Lambda'$ can be identified with $C^\infty(\mathbb{M})$ as follows. Since all elements on $\Gamma(\mathbb{S}')$ can be generated by $\gamma_A$, the generalized Lie derivative on an arbitrary inner product can be written by
\bea
&&\CL_a\left(C\Ket{0'}+\sum_{n=1}^{2D}C_{A_1\cdots A_n}\gamma^{A_1\cdots A_n}\Ket{0'},C'\Ket{0'}+\sum_{m=1}^{2D}C'_{B_1\cdots B_n}\gamma^{B_1\cdots B_m}\Ket{0'}\right)
\cr&=&(\rho(a)C^*C')\braket{0'|0'}+\sum_{m=1}^{2D}(\rho(a)C^*C'_{B_1\cdots B_n})\braket{0'|\gamma^{B_1\cdots B_m}|0'}
+\sum_{n=1}^{2D}(\rho(a)C^*_{A_1\cdots A_n}C')\braket{0'|\gamma^{A_1\cdots A_n}{}^\dagger|0'}
\cr&&+\sum_{n=1}^{2D}\sum_{m=1}^{2D}(\rho(a)C^*_{A_1\cdots A_n}C'_{B_1\cdots B_n})\braket{0'|\gamma^{A_1\cdots A_n}{}^\dagger \gamma^{B_1\cdots B_m}|0'}~.
\eea
Under the identification of $\braket{0'|\gamma_{A_1}\cdots\gamma_{A_n}|0'}$ with constant scalar,
this transformation is equal to the one for $C^\infty(\mathbb{M})$.
Thus, the spin bundle $\Gamma(\mathbb{S}')$ is characterized as the weight $0$ by 
the existence of the inner product $\Gamma(\mathbb{S'})\times\Gamma(\mathbb{S'})\rightarrow C^\infty(\mathbb{M})$.

Then, we identify the dilaton in $\mu$ in a base $\mu_0\in\Gamma(\Lambda)$ as
\be
\mu^\frac{1}{2}=e^{-d}\mu_0^\frac{1}{2}~,
\ee
where $e^{-d}$ is a scalar and the generalized Lie derivative on $\mu_0$ is
\be
\CL_a\mu_0^\frac{1}{2}=\left(\frac{1}{2}\rho(E_A)a^A-\frac{1}{2}\phi'_{BA}{}^Ba^A\right)\mu_0^\frac{1}{2}~.
\ee
In order to recover the dilaton gauge transformation with a weight $\half$, we introduce the gauge transformation of the dilaton $\hat{\delta}_ae^{-d}$ generated by using the above transformation of $\mu^\frac{1}{2}$,
\be
\CL_a\mu^\frac{1}{2}=\CL_a(e^{-d}\mu_0^\frac{1}{2})=(\hat{\delta}_ae^{-d})\mu_0^\frac{1}{2}=\left(-\rho(a)d+\frac{1}{2}\rho(E_A)a^A-\frac{1}{2}\phi'_{BA}{}^Ba^A\right)e^{-d}\mu_0^\frac{1}{2}\ .
\ee
In standard DFT using $\phi'_{AB}{}^C=\Omega^C{}_{AB}$, this relation becomes 
\be
\CL_a\mu^\frac{1}{2}=\CL_a(e^{-d}\mu_0^\frac{1}{2})=(\hat{\delta}_ae^{-d})\mu_0^\frac{1}{2}=\left(-\rho(a)d+\frac{1}{2}\partial_Na^N\right)e^{-d}\mu_0^\frac{1}{2}~.
\ee
Therefore, the gauge transformation $\hat\delta$ of $e^{-d}$ is the one of a half density in the usual sense.

\bibliographystyle{JHEP}
\bibliography{watamura}

\end{document}